\documentclass[]{interact}

\usepackage{epstopdf}% To incorporate .eps illustrations using PDFLaTeX, etc.
\usepackage[caption=false]{subfig}% Support for small, `sub' figures and tables
\usepackage[pagewise]{lineno} % 行番号用（各ページで1から振り直す場合）
\usepackage{hyperref}
\usepackage{lscape}
\usepackage{color}
\usepackage{comment}

\usepackage[numbers,sort&compress]{natbib}% Citation support using natbib.sty
\bibpunct[, ]{[}{]}{,}{n}{,}{,}% Citation support using natbib.sty
\makeatletter% @ becomes a letter
\def\NAT@def@citea{\def\@citea{\NAT@separator}}% Suppress spaces between citations using natbib.sty
\makeatother% @ becomes a symbol again

\theoremstyle{plain}% Theorem-like structures provided by amsthm.sty

\theoremstyle{definition}

\theoremstyle{remark}

\begin{document}
%\linenumbers
\articletype{Research Article}% Specify the article type or omit as appropriate

\title{Multi-task deep neural network for predicting both nuclear fission yields and their experimental errors in peak-shaped data}

\author{
\name{M.~Ueno\textsuperscript{a}\thanks{CONTACT M.~Ueno. Email: ueno@ai.lab.uec.ac.jp}, 
E.~Zhang\textsuperscript{a}, 
K.~Fuchimoto\textsuperscript{b}, 
S.~Chiba\textsuperscript{c}, 
J.~Chen\textsuperscript{d}, 
and C.~Ishizuka\textsuperscript{d}}
\affil{\textsuperscript{a}Graduate School of Informatics and Engineering, The University of Electro-Communications, Tokyo, Japan;}
\affil{\textsuperscript{b}National Center for University Entrance Examinations, Tokyo, Japan;}
\affil{\textsuperscript{c}NAT Research Center, Tokyo, Japan;}
\affil{\textsuperscript{d}Institute of Innovative Research, Institute of Science Tokyo, Tokyo, Japan}
}

\maketitle

\begin{abstract}
The fission product yield (FPY) is crucially important information for numerous nuclear applications. 
However, the peak-shaped characteristics of FPY data present important challenges for predicting unobservable FPY data.
To address these challenges, after applying Multi-task learning models to fission product yield data and their experimental error estimates, we introduce a novel loss function along with incorporation of the odd--even effect. Our approach is intended to predict unknown fission yields and the associated experimental error.
To demonstrate the effectiveness of our proposed method, we compared our proposed method with conventional methods that learn each dataset independently.
Our findings demonstrate that the proposed methods can predict peak-shaped data with experimental error estimates more effectively than earlier methods can.
\end{abstract}

\begin{keywords}
Fission product yield;
Multi-task learning;
Machine learning in nuclear physics;
JENDL-5;
Nuclear fission;
\end{keywords}

\section{\label{sec:introduction}INTRODUCTION}
Nuclear fission data are vitally important in the field of nuclear energy\cite{1bernstein2015nuclear}, including areas of application such as design, simulation, evaluation, and safety. The data provide fundamentally important information for developing nuclear energy technologies. Fission product yield (FPY) data are particularly valuable. However, prominent nuclear data libraries (JENDL\cite{3iwamoto2023japanese}, CENDL\cite{2ge2011updated}, ENDF\cite{4chadwick2011endf}, JEFF\cite{5fission2020fusion}, etc.) have FPY data only for thermal neutron energies with 0.5 and 14 MeV. This situation engenders the need to predict incomplete FPY data at different energies, especially for fast reactors. 

Phenomenological and semimicroscopic models are known to fit existing experimentally obtained data well. Nevertheless, they have poor predictive capability for unobservable FPY data at different energy levels\cite{7schmidt2018review}. Although fully microscopic nuclear fission models have recently been proposed in studies, such as those of Time-Dependent Hartree--Fock--Bogoliubov (TDHFB)\cite{8bulgac2016induced} and Time-Dependent Generator Coordinate Method (TDGCM)\cite{9regnier2016fission}, their computational costs are daunting.
Another theoretical approach to predict FPY is the macroscopic--microscopic (mac-mic) description of nuclear fission. The mac-mic model, which has been developed during more than sixty years, has come to predict FPY using recent models~\cite{5D-langevin,Verriere2021,Albertsson2020,fujio2024} to achieve good agreement with experimentally obtained data. 
Even for those frontier models in nuclear physics, predicting FPY with sufficient accuracy necessary for nuclear data remains very challenging.

Recently, machine learning has come to provide a powerful approach to leverage complex big data for knowledge acquisition and for predictive insights. In the field of nuclear physics, machine learning has become an important tool, with applications in elucidating nuclear structures\cite{10neufcourt2020quantified,11utama2016nuclear,12niu2018nuclear,13wu2020calculation,14utama2016nuclear,15ma2020predictions} and nuclear reactions\cite{16ma2020isotopic,17lovell2020recent,18neudecker2021informing,19wang2021finding}, providing insightful perspectives and approaches to nuclear physics research and demonstrating amazing promise\cite{20bedaque2021ai}.
Among the array of machine learning methods, Bayesian Neural Networks (BNNs), distinguished by their excellent predictive capabilities and their ability to provide uncertainty quantification, are applied frequently for diverse predictive tasks, at which they frequently demonstrate their effectiveness. 
These applications include predicting the charge yields of fission fragments\cite{21qiao2021bayesian} and fission yields\cite{22wang2019bayesian,23wang2021optimizing}, and providing the uncertainty quantification simultaneously.  

Nevertheless, despite the effective predictive capabilities of BNNs, the peak-shaped distribution characteristics of FPY data present important challenges for predicting unobservable FPY data because the FPY data within the peak structure predominantly exhibit a jagged, non-smooth distribution. 
BNNs provide uncertainty quantification by Gaussian-model-based confidence intervals (CIs), which mainly reflect the network estimation uncertainties.
Because a BNN estimates the Gaussian distribution to approximate the data distribution, predict complex structures such as peak structures is difficult. 
%It is difficult for BNN to predict the peak structure accurately because it is not good at learning jagged non-smooth shape data.
%Because of the limited amounts of training data, such complex data interrupt the prediction of unobservable FPY data by BNNs.

As this study focuses primarily on improving the predictive performance for the peak-shaped FPY distribution, we do not pursue the estimation of CIs in this work. 
Instead, this study addresses the FPY error, which represents another type of error distinct from the network estimation uncertainty. 
The Japanese Evaluated Nuclear Data Library ver. 5 (JENDL-5) \cite{3iwamoto2023japanese} provides not only FPY value but also the associated FPY error \cite{FPY_covariance}, which is estimated via error propagation to obtain the experimental error.
In the JENDL evaluation, the uncertainties of fission product yields are estimated using a generalized least-squares method. 
Specifically, the covariance matrix of the evaluated yields is estimated so as to reproduce the standard deviations of the mass-yield distribution given in the well-established England–Rider systematics. 
The FPY error values are obtained as the square roots of the diagonal elements of this covariance matrix.

These facts indicate that FPY and its FPY error are strongly correlated, with information from one exerting a significant influence on the other. 
Consequently, in general, nuclides with higher FPY values tend to have lower relative FPY errors due to this strong correlation. 
This correlation prominently reflects the characteristics of the relevant fission product nuclides, much in the same way that the correlation between nuclear mass and deformation represents underlying nuclear properties.
Multi-task models are known to enable supplementary prediction of multiple different mutually related types of data while assisting their learning each other. Recently, the Multi-task model has been used to learn multiple physics observations simultaneously, such as nuclear masses and deformations \cite{bai2021description,wu2022multi}, as well as nuclear number densities and corresponding binding energies for various nuclear shapes \cite{hizawa2024nonempirical}. 
Similarly, predicting both FPY value and the associated FPY error simultaneously using a Multi-task model might be more effective than using conventional methods to predict each dataset independently.

To improve the Multi-task model prediction accuracy, Multi-gate Mixture-of-Experts (MMoE)\cite{ma2018modeling} has been proposed.
After MMoE structurally learns shared representations from input data through multiple feedforward neural networks, it allocates task-specific sub-networks to learn task-dependent features.
In fact, because of its efficient information-sharing capabilities, MMoE has demonstrated superior results for Multi-task prediction across various fields such as aerospace\cite{aero-engine}, industrial fault detection\cite{industrial_detection}, and traffic data analysis\cite{traffic_data_analysis}.

Inspired by this important benefit, this study proposes using the MMoE to learn and predict FPY and FPY errors simultaneously.
The proposed method is expected to increase the prediction accuracies of both FPY and FPY error data because MMoE is known to be effective when the multiple tasks affect one another.
A key advantage of MMoE in contrast to BNN, is its superior capability in predicting complex structures like peak data.

Furthermore, to predict peak data accurately, we introduce a novel loss function that assigns weights to the training data's loss function according to the FPY values. 
Additionally, we incorporate the odd--even effect\cite{25schmidt2018review} as auxiliary input information. 

The MMoE framework is introduced because FPY and FPY error are strongly correlated targets, and joint learning can exploit this correlation to improve both predictions. 
In contrast, the weighted loss function is introduced to address the underfitting of the peak region caused by standard MSE-based training, thereby improving the reproduction of the peak-shaped FPY distribution.

The results demonstrate that the proposed methods can predict the peak-shaped FPY distribution with experimental error estimates more effectively.

\section{\label{sec:theoretical-framework}THEORETICAL FRAMEWORK}
\subsection{FPY and FPY error data}
This study utilizes the fission product yields (FPYs) and their associated errors data from JENDL-5. 
Here, the term “FPY error” refers to the evaluated uncertainty assigned to each independent fission product yield in the JENDL evaluation \cite{FPY_covariance}, which is derived through a generalized least-squares (GLS) procedure under five physical constraints, which are conservation of mass number, conservation of charge number, normalization of the independent yields, normalization of the heavier-mass yields, and consistency with calculated mass-chain yields.
Specifically, the updated yields $\boldsymbol{\theta}_{\mathrm{upd}}$ and their covariance matrix $\mathbf{V}_{\mathrm{upd}}$ are obtained by the GLS method as
\begin{align}
\boldsymbol{\theta}_{\mathrm{upd}} &= \boldsymbol{\theta}_\mathrm{a} 
+ \mathbf{V}_\mathrm{a} \mathbf{S}^{\top}
\left( \mathbf{S} \mathbf{V}_\mathrm{a} \mathbf{S}^{\top} + \mathbf{V} \right)^{-1}
(\boldsymbol{\eta} - \mathbf{S}\boldsymbol{\theta}_\mathrm{a}), 
\\
\mathbf{V}_{\mathrm{upd}} &= \mathbf{V}_\mathrm{a} 
- \mathbf{V}_\mathrm{a} \mathbf{S}^{\top}
\left( \mathbf{S} \mathbf{V}_\mathrm{a} \mathbf{S}^{\top} + \mathbf{V} \right)^{-1}
\mathbf{S} \mathbf{V}_\mathrm{a},
\end{align}
where $\boldsymbol{\theta}_\mathrm{a}$ denotes the initial FPY vector, $\mathbf{S}$ denotes the sensitivity matrix associated with the imposed physical constraints, $\boldsymbol{\eta}$ denotes the vector of constraint values, and $\mathbf{V}$ denotes the uncertainty associated with the constraint quantities in the GLS formulation.
The initial covariance matrix $\mathbf{V}_\mathrm{a}$ was taken from experimental uncertainties of the independent yields when available. 
When experimental uncertainties were not available, errors estimated from JENDL/FPY-2011 or the JEFF-3.3/FY library, incorporating improvements from the UKFY3.7 evaluation \cite{mills2017new}, were adopted as pseudo-experimental data.

Through this GLS adjustment, the evaluated FPY values and their covariance matrix become mutually consistent with conservation laws and experimental information. 
The diagonal elements of $\mathbf{V}_{\mathrm{upd}}$ correspond to the variances of FPY, and the square roots of these diagonal elements are defined as the “FPY errors” in this study.

It should be emphasized that the FPY errors represent evaluated experimental uncertainties propagated and adjusted under physical constraints, rather than statistical fluctuations from a Poisson counting process. 
Although experimental fission yield measurements originate from counting statistics that are often approximated by Poisson distributions, the final FPY errors in JENDL are obtained after global adjustment and covariance propagation using the GLS method. 
Therefore, they reflect correlated uncertainties among nuclides, rather than independent counting errors.

This study proposes a multi-task neural network to simultaneously predict both the evaluated FPY values and these associated evaluated uncertainties.

\subsection{\label{subsc:MMoE}MMoE for predicting both FPY and FPY error data}
To learn and predict FPY and FPY error data, this study employs the MMoE model \cite{ma2018modeling} as the core feature of the multi-task learning framework. The MMoE is built on the fundamental concept of assembling a set of feedforward neural networks for the collective learning of complex patterns and representations from data. At its core, MMoE has a gating mechanism that assigns weights to each feedforward neural network for a given input instance. These weights are learned to combine the multiple feedforward neural networks and thereby to increase prediction accuracy.

The MMoE is given as
\begin{align}
\hat{y}^k &= h^k (f^k (\bm{x})),
\end{align}
where $\hat{y}^k$ represents the output of the $k$-th task, $\bm{x} = (x_1, x_2, \ldots, x_n)$ represents the input vector, and $n$ stands for the number of the input data, whereas $h^k$ and $f^k(\bm{x})$ denote the feedforward neural networks for the $k$-th task.

\begin{align}
f^k (\bm{x})= \sum_{j=1}^{m}g^k_j (\bm{x}) f_j (\bm{x})\text{, and}
\end{align}

\begin{align}
g^k_j (\bm{x})
=
\frac{\exp\!\left( \bm{W}_{g^k_j}\bm{x} \right)}
{\sum_{j^{\prime}=1}^{m} \exp\!\left( \bm{W}_{g^k_{j^{\prime}}}\bm{x} \right)}
\end{align}

\begin{figure}[tb]
\centering
\includegraphics[width=0.47\textwidth]{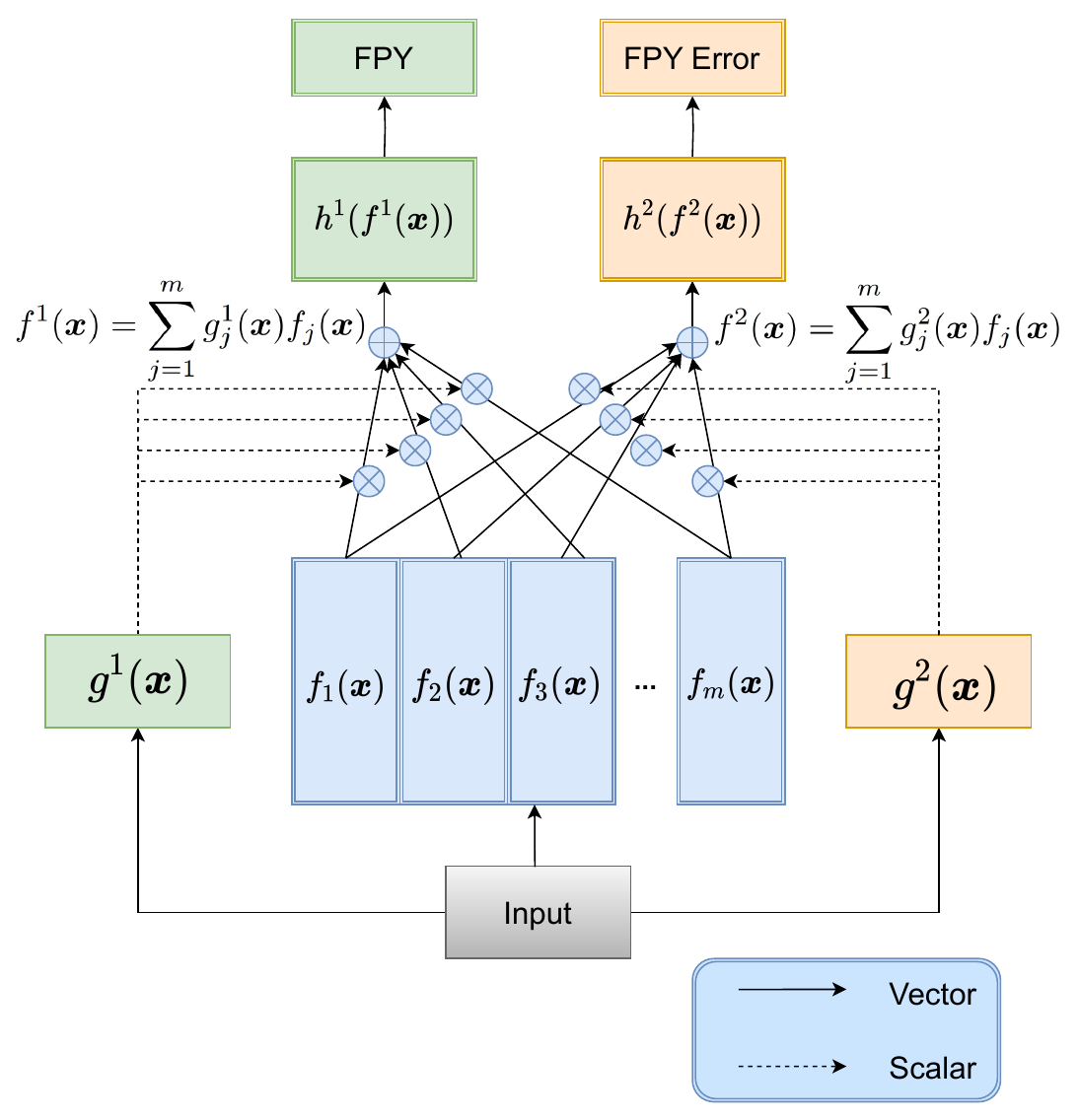}
\caption{Network architecture of the proposed method.}
\label{fig:1}
\end{figure}

Each task corresponds to a task gate $g^k_j (\bm{x})$, where each task gate is a softmax layer. In addition, $m$, which denotes the number of feedforward neural networks, is set as 3. Here, $g^k_j(\bm{x})$ represents the corresponding weight of the $j$-th feedforward neural network given by the gate network. 
In addition, $f_j (\bm{x})$ represents $j$-th feedforward neural network. The purpose of the gate network for each task is to select and weight all the feedforward neural networks. Within this framework, $\bm{W}_{{g}^{k}_j}$ is the trainable matrix. Different tasks can match their respective loss functions and weights. Additional details about MMoE are presented in the relevant literature \cite{ma2018modeling}.

\autoref{fig:1} depicts the network architecture of the proposed MMoE. The proposed neural network has two outputs: FPY value as $\hat{y}^1$ and its FPY error as $\hat{y}^2$. 
This architecture enables the simultaneous prediction of the FPY values and their FPY errors using supervised learning from observed data from JENDL-5. 
For this study, we specifically examine the mass distribution of the fission products, FPY($A$), because the available data are abundant compared to the independent yields, FPY($Z, A, m$), where $Z$ stands for the charge number, $A$ expresses the mass number of the fission fragment, $m$ denotes the isomeric state.
The architecture shown in \autoref{fig:1} enables learning of FPY data and learning their FPY errors supplementarily while assisting each other.  

The overall loss function for training the proposed model comprises two terms: the loss function $Loss_{FPY}$ for the FPY prediction task and the loss function $Loss_{ERROR}$ for the FPY error prediction task.

\begin{align}
L_{\text{total}} &= \alpha * Loss_{FPY} + (1 - \alpha) * Loss_{ERROR}.
\end{align}
Parameter $\alpha$ is a tuning parameter that adjusts the weight balance between the FPY prediction and the FPY error prediction. 
We determined $\alpha=0.9$ as its optimal value through grid search over the set $\{0.5, 0.6, 0.7, 0.8, 0.9\}$ to maximize only the FPY prediction because FPY prediction is more important than the FPY error prediction.

The first problem is which loss functions this study employs. The most popular loss function is the following mean squared error (MSE),
% \begin{align}
% MSE &= \frac{1}{n} \sum_{i=1}^{n} L_i^{\text{M}}, \\
% L_i^{\text{M}} &= (y_i - \hat{y}^{\text{M}}_i)^2.
% \end{align}
\begin{align}
\text{MSE}^k &= \frac{1}{n} \sum_{i=1}^n \left( y_{i}^k - \hat{y}_{i}^k \right)^2,
\end{align}
where $k=1$ signifies the FPY prediction task and where $k$=2 signifies the FPY error prediction task.
Also, $y^k_i$ represents the true value of the $i$-th data point for task $k$; $\hat{y}^k_i$ stands for the predicted value of the $i$-th data point for $k$-th task.
That is to say, the ${MSE}^k$ (5) signifies MSE of $k$-th task for the dataset (the sample size $n$) within each batch.
Although the important challenge of this study is the prediction of the peak data of FPY data, as described previously, the MSE function is unable to catch the peak data characteristics sufficiently.

Next, we propose an innovative loss function for FPY data to predict the peak data effectively, instead of $MSE^{1}$.
The proposed loss function, which we designate as a "weighted loss function", is designed to allow the model to examine peak data specifically, ultimately improving the predictive capacity.
It is noteworthy that the proposed loss function is only used for the prediction of FPY, not for prediction of FPY error. 
The proposed "weighted loss function" is defined as presented below.

\begin{align}
weightedLoss= \frac{1}{n} \sum_{i=1}^{n}L_i^{\text{W}},
\end{align}
where
\begin{align}
    L_i^{\text{W}} &= 
    \begin{cases}
      \beta * (y_i  - \hat{y}_i^{\text{W}})^2 ,& normalize(y_i)< r \\
     w(y_i) * (y_i  - \hat{y}_i^{\text{W}})^2 ,& normalize(y_i)\geq r\\
    \end{cases},\label{eq:2}
\end{align}

\begin{align}
\text{and}\quad
w(y_i)=1+normalize(y_i) .
\label{eq:9}
\end{align}

The $weightedLoss$ (6) represents the total loss value of the FPY prediction for the dataset (of sample size $n$) within each batch. 
Also, $L_i^{\text{W}}$ denotes the prediction error for the $i$-th data point within a batch, as obtained using the proposed method, where $y_i$ represents the true value of the $i$-th data point. Also, $\hat{y}_i^{\text{W}}$ represents the proposed method's prediction result for the $i$-th data point. 
Here, superscript \( \mathrm{W} \) denotes that the value is obtained from the weighted loss function.
To prevent an excessively large grid search over $(\alpha,\beta,r)$ in the proposed multi-task model, the tuning parameters $(\beta,r)$ of the weighted loss were determined in advance using a single-task deep neural network (DNN) with $\alpha=1$, where only the FPY prediction term is optimized.
Specifically, $(\beta, r)$ is selected from 25 possible combinations, defined as the Cartesian product $(\beta, r) \in \{0.001, 0.005, 0.01, 0.02, 0.05\} \times \{0.5, 0.6, 0.7, 0.8, 0.9\}$, where $\beta$ and $r$ are tuning parameters.
Therefore, the selected $(\beta,r)$ are then fixed when training the proposed multi-task model.
%Because the weighted loss function is designed to predict only the FPY value, the tuning is conducted with a (single-task) deep neural network (DNN).
Also, $w(y_i^{\text{W}})$ stands for the weight coefficients; $normalize(y_i)$ denotes the standardized score of $y_i$ with mean 0.0 and standard deviation 1.0.
In \autoref{eq:9}, $w(y_i)$ functions to assign higher weights to larger normalized values of $y_i$ to force the model to fit a larger value data of FPY better.

When $normalize(y_i)$ is larger than tuning parameter $r$,
the input vector $\bm{x_i}$ is called "peak data".
When $normalize(y_i)$ is smaller than the tuning parameter $r$,
the input vector $\bm{x_i}$ is designated as "non-peak data".
By assigning an extremely small value to $\beta$, the weight of non-peak data is diminished. It increases the model's learning sensitivity to peak data. Additionally, optimizing the value of $w(y_i)$ adjusts the peak data weight according to the training data FPY value. Consequently, as the FPY value of the training data increases, the corresponding weight coefficients also increase. This strategic weighting forces the model learning to prioritize the prediction performance of the peak data.

\section{\label{sec:RESULTS AND DISCUSSION}RESULTS AND DISCUSSION}
\subsection{Comparison of prediction errors}
\label{subsec:comparison-bnn-dnn}
First, to evaluate the peak data prediction accuracies of the Multi-task DNN methods, we conducted comparisons using the earlier BNN method and a (single-task) deep neural network method.

\subsubsection{Datasets}
\label{subsubsec:dataset}
To evaluate the effectiveness of the proposed methods, we use four distinct sets of training and testing data from JENDL-5 and ensure that the data in the test set are not included in the training set, as described below

Data Set 1: 
Training Set: $^{236,237,238}$U, $^{242,244,245,246,248}$Cm, $^{254,255,256}$Fm, $^{239,241,242}$Pu, $^{249,250,252}$Cf, $^{241,242,243}$Am, $^{237,238}$Np, $^{231}$Pa, $^{253,254}$Es, $^{227,229,232}$Th, and $^{254,255,256}$Fm.
Testing Set: $^{235}$U.

Data Set 2: Training Set: $^{235,236,237}$U, $^{242,244,245,246,248}$Cm, $^{254,255,256}$Fm, $^{239,241,242}$Pu, $^{249,250,252}$Cf, $^{241,242,243}$Am, $^{237,238}$Np, $^{231}$Pa, $^{253,254}$Es, $^{227,229,232}$Th, and $^{254,255,256}$Fm.
Testing Set: $^{238}$U.

Data Set 3: Training Set: $^{233,234,235,236,237,238}$U, $^{242,244,245,246,248}$Cm, $^{254,255,256}$Fm, $^{238,240,241,242}$Pu, $^{249,250,252}$Cf, $^{241,242,243}$Am, $^{237,238}$Np, $^{231}$Pa, $^{253,254}$Es, $^{227,229,232}$Th, and $^{254,255,256}$Fm.
Testing Set: $^{239}$Pu.

Data Set 4: Training Set: $^{233,234,235,236,237,238}$U, $^{242,244,245,246,248}$Cm, $^{254,255,256}$Fm, $^{238,239,240,242}$Pu, $^{249,250,252}$Cf, $^{241,242,243}$Am, $^{237,238}$Np, $^{231}$Pa, $^{253,254}$Es, $^{227,229,232}$Th, and $^{254,255,256}$Fm.
Testing Set: $^{241}$Pu.

The testing sets $^{235}$U, $^{238}$U, $^{239}$Pu, and $^{241}$Pu were selected because they are major actinides in the JENDL library for which abundant and reliable experimental FPY data are available. 
These nuclides are also of primary importance in practical reactor analyses, including light-water and fast reactors, and therefore serve as representative benchmarks for evaluating predictive performance. 
In contrast, other nuclides in JENDL often have more limited experimental information, and the predictive accuracy for such nuclides may be affected by insufficient training data. 

To examine the influence of data augmentation (DA) on the prediction performance, two control settings were prepared: one using DA and the other without DA.
In the case without DA, the training dataset contained 6,039 samples in total. These samples were composed of 80\% of the JENDL-5 FPY entries (5,382 data points) together with additional experimental datasets consisting of 286~\cite{peak1gooden2016energy}, 185~\cite{NAIK2013185}, and 189~\cite{NAIK201516} data points taken from previous studies. The remaining 20\% of the JENDL-5 FPY data (1,071 samples) were reserved for validation.
When the DA procedure was applied, the number of training samples increased to 16,713, whereas the validation dataset was kept identical to that used in the non-DA case.

\subsubsection{Experimental settings}
\label{subsubsec:settings}
For the experiments, the network input vector $\bm{x_i}$ comprises four values $\bm{x_i}=(Z_i, N_i, A_i, E_i)$, where $Z_i$ signifies the charge number, $N_i$ represents the neutron number of the fission nucleus, $A_i$ represents the mass number of the fission fragment, and $E_i$ denotes the excitation energy of the compound nucleus.

% \afterpage{\FloatBarrier}

% 7 layersăžă§ĺčĄç çŠśă§ăŻć˘ăŤĺŽé¨ăăŚEăăEăŻăźăŻăćˇąăăăăă¨ăŤĺŠçšăăŞăEă¨ăĺEćć¸ăż
%Generally, with similar number of connection weights, the shallow network would be more dependent on the prior input. In contrast, the deep network would be more dependent on its deduction capability. Therefore, for specific problems, there should be a balanced choice of network structure. We also tested network structures of 11-12-12, 9-10-10-10, 9-9-8-8-9, 8-8-8-8-7-7, 7-7-7-7-7-7-8 neurons for 3, 4, 5, 6, 7 hidden layers, respectively. Note that all these structures have similar number of connection weights to that of the 16-16 structure. Correspondingly, the total ĎE2 N are 4.07ĂE0âE (3 layers), 4.99ĂE0âE (4 layers), 4.64ĂE0âE (5 layers), 5.05ĂE0âE (6 layers), 4.94ĂE0âE (7 layers). We see that deep networks have no advantages in this work. The best network is the double-layer structure of 16-16 neurons. In addition, the deep networks take much longer computing time to get convergence.

For BNNs, FPY prediction requires marginalization over the posterior probability distribution of the network parameters.
This marginalization involves an intractable integral that cannot be evaluated analytically.
Therefore, we employ the Markov Chain Monte Carlo (MCMC) method to approximate the posterior probability distribution, according to reports of earlier study \cite{23wang2021optimizing}. 
By sampling from the posterior distribution, the MCMC-based BNN enables not only point prediction of FPY values but also uncertainty quantification, such as the estimation of confidence intervals for the predicted FPYs.

For learning BNNs, we employ the MCMC method with two hidden layers, each containing 16 neurons, according to reports of earlier study \cite{23wang2021optimizing}. 
In this study, the MCMC-based BNN is implemented in NumPyro library, which is probabilistic programming framework.
Posterior inference was performed using the No-U-Turn Sampler (NUTS) \cite{hoffman2014no}, an adaptive variant of Hamiltonian Monte Carlo (HMC) \cite{duane1987hybrid}, as implemented in NumPyro.
After discarding the burn-in samples, posterior samples of the network parameters are used to compute the predictive mean and confidence intervals of FPY.

%For learning BNNs, we employ the MCMC method with two hidden layers, each containing 16 neurons, according to reports of earlier study \cite{23wang2021optimizing}. 
This configuration was justified because 
Wang et al. \cite{23wang2021optimizing} demonstrated that increasing the number of hidden layers up to seven did not improve the prediction performance, and demonstrated that the best results were obtained with a shallow architecture consisting of two hidden layers, each with neurons.
Additionally, BNN is unable to incorporate the weighted loss function (7) because it employs not the minimum loss function estimates but the MCMC estimates.

For learning DNNs, we construct 16 neurons per layer, as with BNN. The number of hidden layer 10 was ascertained by grid search from $\{5, 10, 15 \}$.

For learning Multi-task DNNs, we employ 16 neurons per layer and 10 hidden layers, as with DNNs for control under the same experiment conditions. 
Consequently, the Multi-task DNN includes both $f_j (\bm{x})$ and $h^k (\bm{x})$, each comprising five hidden layers with 16 neurons per layer. 
In this study, the number of experts was selected by grid search over ($m=1,2,3,4, 5$), using the validation performance of FPY prediction as the selection criterion. 
Among these candidates, ($m = 3$) provided the best overall validation performance while avoiding unnecessary model complexity.
%Additionally, we ascertained the number of experts as 3 using a grid search from 1 to 5.
% The loss function of the Multi-task DNN is the MSE. For the FPY prediction task, the weight of loss function, which is a tuning parameter, is set as 0.9. For the FPY error prediction task, the weight of loss function is set to 0.1. The set of weights for the loss functions is optimized by grid search from \{0.6/0.4, 0.7/0.3, 0.8/0.2, 0.9/0.1\}.

To improve the accuracy of FPY prediction further, we incorporate the odd--even effect into the input.
In the mass distribution of fission products, the odd--even effect is a prominent phenomenon that originates from the nucleon pairing effect and from differences in nuclear binding energy. 
During the fission process, nuclei with even mass numbers are generally more stable and exhibit higher yields than those with odd mass numbers, producing a characteristic jagged pattern in the fission product yield distribution \cite{25schmidt2018review}. 
Reports of recent studies \cite{22wang2019bayesian,23wang2021optimizing} have described that the odd--even effect helps to capture the local variations in the yield distribution better and improves the predictive accuracy of FPY models.
To account for the odd–even effect, we specifically introduce the parity indicator $O_i$ as an additional input feature as follows:
when $A_i$ is even, $O_i=1$, representing the even-mass enhancement effect; when $A_i$ is odd, $O_i=0$. 
For non-peak data, the value of $O_i$ is set to 0.5 in order to reduce the parity contrast in regions where the odd--even effect is less pronounced.
It is noteworthy that the value of variable $O_i$ is not a physical observable but a binary encoding of the mass-number $A_i$ parity.

As described in Section \ref{subsc:MMoE}, the classification of peak and non-peak data is determined according to \autoref{eq:2} based on the normalized FPY value. 
Specifically, data points satisfying $normalize(y_i) \ge r$ are classified as peak data, whereas those with $normalize(y_i) < r$ are classified as non-peak data. 
Accordingly, the value of $O_i$ is assigned automatically based on this classification.
%For odd--even effects, $O_i$ is added to the input vector as the odd--even value. When the mass number $A_i$ is even, $O_i=1$; when the $A_i$ is odd, $O_i=0$. To mitigate the adverse influence of the odd--even effect on non-peak data, only the $O_i$ values for peak data are set as 0 or 1, whereas the $O_i$ values for non-peak data are set uniformly to 0.5.

\subsubsection{Performance comparison among BNN, DNN and the proposed method}
\label{subsubsec:results}
We compare the prediction errors of the FPY values and the FPY error values among BNN, DNN and Multi-task DNN for both peak and non-peak data.
For this study, we use data from JENDL-5 \cite{3iwamoto2023japanese}, which includes both evaluated FPY values and FPY error data. 
Particularly, the FPY error is estimated via error propagation to obtain the experimental error, considering the covariance matrix of the variables necessary for estimating the FPY value \cite{FPY_covariance}.
The FPY and the associated FPY error in JENDL are evaluated based on equations (6) and (7) in \cite{FPY_covariance}, respectively.
Our multi-task learning method learns these evaluation results as training data.  
Equations (6) and (7) share a common second term. 
In this case, the proposed method is expected to increase the prediction accuracies of both the FPY data and the FPY error because the architecture enables learning the FPY and learning the FPY error supplementarily, with mutual assistance. 
To ensure this advantage, we train the BNN and the DNN using the FPY data and the FPY error independently to compare their predictions with those of the Multi-task DNN. 
Particularly, this advantage can be evaluated directly by comparing the performances of the proposed method and (single task) DNNs, which learn the FPY data and the FPY error data independently, because it can be controlled under the same experiment conditions.
Specifically, we compare the prediction performances of BNN, DNN using the MSE loss function ("DNN (MSE)"), DNN using the weighted loss function ("DNN (Weighted loss function)"), DNN using the MSE loss function with odd--even values ("DNN (MSE) + odd--even"), DNN using the weighted loss function with odd--even values ("DNN (Weighted loss function) + odd--even"), Multi-task DNN using the MSE ("Multi-task DNN (MSE)"), Multi-task DNN using the weighted loss function ("Multi-task DNN (Weighted loss function)"), Multi-task DNN using the MSE with odd--even values ("Multi-task DNN (MSE) + odd--even"), and Multi-task DNN using the weighted loss function with odd--even values ("Multi-task DNN (Weighted loss function) + odd--even").
Because a BNN estimates the Gaussian distribution to approximate the data distribution, predictive evaluation based only on the expected value estimates of FPY without using CIs might impede the BNN. Therefore, it is noteworthy that this experiment is limited to evaluating point estimates related to prediction of the detailed peak structure.
\autoref{tab:1} presents the validation errors for both FPY and FPY error for each method at energies of $2.53 \times 10^{-8}$ MeV (thermal neutron energy), 0.5 MeV, and 14 MeV.

In \autoref{tab:1}, "energy" represents the excitation energy in MeV. The "average total $\chi^2$", "average peak $\chi^2$", and "average non-peak $\chi^2$" respectively stand for the mean values obtained under various experiment conditions for total $\chi^2$, peak $\chi^2$, and non-peak $\chi^2$. Total $\chi^2$ is defined as $\sum_{i}[t_i-f(x_i)]^2/n$, where $t_i$ signifies the $i$-th ground truth value of the test set.
The error $\chi^{2}$ value ("peak $\chi^2$") corresponds only to the peak structure, whereas the error $\chi^{2}$ value ("non-peak $\chi^2$") corresponds only to the non-peak structure. 

Similarly, the "FPY error average total $\chi^2$", "FPY error average peak $\chi^2$", and "FPY error average non-peak $\chi^2$" respectively represent the mean values obtained under various experiment conditions for FPY error total $\chi^2$, FPY error peak $\chi^2$, and FPY error non-peak $\chi^2$. 
Here, FPY error total $\chi^2$ represents the total FPY error value. FPY error peak $\chi^2$ denotes the FPY error value for the peak structure. 
In addition, FPY error non-peak $\chi^2$ stands for the FPY error value for the non-peak structure.
\clearpage

\begin{landscape}
\begin{table*}[t]%The best place to locate the table environment is directly after its first reference in the text   [!htbp]
\caption{Average validation errors $\chi^{2} (10^{-6})$ of various models under three energy conditions}
\footnotesize
\label{tab:1}
\begin{tabular}{lccccccc}
&
&
\multicolumn{3}{c}{\textrm{average}}&
% \multicolumn{1}{l}{\textrm{average}}&
% \multicolumn{1}{l}{\textrm{average}}&
% \multicolumn{1}{l}{\textrm{average}}&
% \multicolumn{1}{l}{\textrm{average}}&
%\multicolumn{3}{c}{\textrm{uncertainty}}\\
\multicolumn{3}{c}{\textrm{FPY error}}\\

&
&
\multicolumn{1}{l}{\textrm{  total}}&
\multicolumn{1}{l}{\textrm{  peak}}&
\multicolumn{1}{l}{\textrm{non-peak}}&
\multicolumn{1}{l}{\textrm{average}}&
\multicolumn{1}{l}{\textrm{average}}&
\multicolumn{1}{l}{\textrm{average}}\\
\textrm{Methods}&
\textrm{energy}&
\textrm{$\chi^{2}$}&
\textrm{$\chi^{2}$}&
\textrm{$\chi^{2}$}&
% \multicolumn{1}{l}{\textrm{$\chi^{2}$}}&
% \multicolumn{1}{l}{\textrm{$\chi^{2}$}}&
\textrm{total $\chi^{2}$}&
\textrm{peak $\chi^{2}$}&
\multicolumn{1}{l}{\textrm{non-peak $\chi^{2}$}}\\
\hline
% Pu241 0.00025
BNN  &$2.53 \times 10^{-8}$ & 4.66 & 6.74 & 4.03 & 10.45 & 33.79 & 3.34\\
DNN (MSE) & $2.53 \times 10^{-8}$ & 7.09 & 8.30 & 6.73 & 7.47 & 19.14 & 3.91\\
DNN (Weighted loss function) & $2.53 \times 10^{-8}$ & 9.51 & 7.18 & 10.22 & 7.47 & 19.14 & 3.91\\
DNN (MSE) + odd--even & $2.53 \times 10^{-8}$ & 4.58 & 9.36 & 3.13 & 6.38 & 18.59 & 2.65\\
DNN (Weighted loss function)+oddeven & $2.53 \times 10^{-8}$ & 4.52 & 9.26 & \textbf{3.08} & 6.38 &18.59 & 2.65\\
Multi-task DNN (MSE)  &$2.53 \times 10^{-8}$ & 4.46 & 6.28 & 3.90 & 6.81 & 17.47 & 3.56\\
Multi-task DNN (Weighted loss function)  &$2.53 \times 10^{-8}$ & 5.00 & 5.31 & 4.90 & 7.05 & 18.82 & 3.46\\
Multi-task DNN (MSE) + odd--even  &$2.53 \times 10^{-8}$ & 4.43 & 5.53 & 5.87 & 4.20 & 11.36 & 2.01\\
Multi-task DNN (Weighted loss function) + odd--even  &$2.53 \times 10^{-8}$ & \textbf{3.36} & \textbf{4.26} & \textbf{3.08} & \textbf{3.75} & \textbf{9.87} & \textbf{1.89}\\
\hline
BNN  & 0.5& 7.49 & 15.70 & 4.99  & 9.13 & 29.5 & 2.92\\
DNN (MSE) & 0.5 & 7.19 & 15.53 & 4.80 & 4.70 & \textbf{9.44} & 3.10\\
DNN (Weighted loss function) & 0.5 & 8.26 & 14.52 & 6.50 & 4.70 & \textbf{9.44} & 3.10\\
DNN (MSE) + odd--even & 0.5 & 6.53 & 16.59 & 2.97 & 4.41 & 10.48 & 2.56\\
DNN (Weighted loss function) + odd--even & 0.5 & 5.95 & 13.97 & 3.65 & 4.41 & 10.48 & 2.56\\
Multi-task DNN (MSE)  &0.5 & 6.94 & 17.65 & 3.67 & 4.06 & 12.25 & 1.23\\
Multi-task DNN (Weighted loss function)  &0.5 & 6.26 & 13.98 & 3.91 & 5.05 & 15.40 & 1.89\\
Multi-task DNN (MSE) + odd--even  &0.5 & 6.53 & 14.44 & 4.12 & \textbf{3.16} & 9.57 & 1.20\\
Multi-task DNN (Weighted loss function) + odd--even  &0.5 & \textbf{5.50} & \textbf{11.60} & \textbf{3.63} & 3.21 & 10.10 & \textbf{1.11}\\
\hline
BNN  &14 & 15.03 & 39.90 & 7.45 & 9.10 & 17.06 & 6.67\\
DNN (MSE) & 14 & 10.02 & 22.17 & 6.98 & 4.99 & 8.07 & 4.06\\
DNN (Weighted loss function) & 14 & 11.02 & 19.62 & 8.41 & 4.99 & 8.07 & 4.06\\
DNN (MSE) + odd--even & 14 & 8.74 & 17.58 & 6.05 & 6.37 & 12.14 & 4.41\\
DNN (Weighted loss function) + odd--even & 14 & 8.24 & 15.65 & 5.97 & 6.37 & 12.14 & 4.41\\
Multi-task DNN (MSE)  &14 & 9.57 & 21.05 & 6.07 & 3.50 & \textbf{5.27} & 2.95\\
Multi-task DNN (Weighted loss function)  &14 & 8.06 & 20.77 & \textbf{4.19} & 2.30 & 5.88 & 3.82\\
Multi-task DNN (MSE) + odd--even  &14 & 9.77 & 23.27 & 5.65 & \textbf{2.12} & 5.76 & \textbf{1.01}\\
Multi-task DNN (Weighted loss function) + odd--even  &14 & \textbf{7.35} & \textbf{13.08} & 5.60 & 2.48 & 5.59 & 1.53\\ \hline
\end{tabular}
%\end{minipage}
%\hfill
\end{table*}
\end{landscape}
%\FloatBarrier

%ĂĽÂÂĂ¨ÂĄÂ¨ĂŚÂ Âź
\begin{landscape}
\begin{table*}[t]%The best place to locate the table environment is directly after its first reference in the text[!htbp]
\caption{Validation errors $\chi^{2} (10^{-6})$ for $^{235}$U obtained using various methods}
\begin{minipage}{0.4\linewidth}
\footnotesize
\label{tab:2}
\begin{tabular}{lcccccccc}
&
&
&
\multicolumn{1}{l}{\textrm{total}}&
\multicolumn{1}{l}{\textrm{peak}}&
\multicolumn{1}{l}{\textrm{non-peak}}&
%\multicolumn{3}{c}{\textrm{uncertainty}}\\
\multicolumn{3}{c}{\textrm{FPY error}}\\
% \multicolumn{1}{l}{\textrm{uncertainty}}&
% \multicolumn{1}{l}{\textrm{uncertainty}}\\
\textrm{Methods}&
\textrm{nuclei}&
\textrm{energy}&
\textrm{$\chi^{2}$}&
\textrm{$\chi^{2}$}&
\textrm{$\chi^{2}$}&
% \multicolumn{1}{l}{\textrm{$\chi^{2}$}}&
% \multicolumn{1}{l}{\textrm{$\chi^{2}$}}&
\textrm{total $\chi^{2}$}&
\textrm{peak $\chi^{2}$}&
\multicolumn{1}{l}{\textrm{non-peak $\chi^{2}$}}\\
\hline
%U235
BNN & $^{235}$U &0.5& 9.33 & 24.30 & 4.77 & 6.30 & 16.20 & 3.28\\
DNN (MSE) & $^{235}$U & 0.5 & 6.53 & 18.49 & 2.89 & 3.01 & \textbf{8.54} & 1.32\\
DNN (Weighted loss function) & $^{235}$U & 0.5 & 8.40 & 18.24 & 5.39 & 3.01 & \textbf{8.54} & 1.32\\
DNN (MSE) + oddeven & $^{235}$U & 0.5 & 8.09 & 24.62 & 3.05 & 4.75 & 16.60 & 1.14\\
DNN (Weighted loss function) + oddeven & $^{235}$U & 0.5 & 6.64 & 19.59 & 2.69 & 4.75 & 16.60 & 1.14\\
Multi-task DNN (MSE) & $^{235}$U &0.5& 7.30 & 17.85 & 4.09 & \textbf{2.69} & 9.19 & 0.76\\
Multi-task DNN (Weighted loss function) & $^{235}$U &0.5& 5.91 & 16.67 & 2.63 & 3.50 & 12.56 & 0.74\\
Multi-task DNN (MSE) + odd--even & $^{235}$U &0.5& 7.00 & 17.16 & 3.90 & 2.77 & 9.98 & \textbf{0.57}\\
Multi-task DNN (Weighted loss function) + odd--even & $^{235}$U &0.5& \textbf{5.71} & \textbf{16.58} & \textbf{2.39} & 4.05 & 15.12 & 0.68\\
\hline
BNN & $^{235}$U &14 & 20.70 & 62.96 & 7.81 & 6.83 & 12.75 & 5.03\\
DNN (MSE) & $^{235}$U & 14 & 5.66 & 14.02 & 3.12 & 2.82 & 6.46 & 1.71\\
DNN (Weighted loss function) & $^{235}$U & 14 & 5.46 & 11.73 & 3.56 & 2.82 & 6.46 & 1.71\\
DNN (MSE) + odd--even & $^{235}$U & 14 & 5.47 & 12.75 & 3.25 & 6.38 & 18.59 & 2.65\\
DNN (Weighted loss function) + odd--even & $^{235}$U & 14 & 6.15 & 12.04 & 4.35 & 6.38 & 18.59 & 2.65\\
Multi-task DNN (MSE) & $^{235}$U &14 & 8.91 & 20.08 & 5.50 & 2.37 & 6.68 & \textbf{1.05}\\
Multi-task DNN (Weighted loss function) & $^{235}$U &14 & 5.96 & 11.44 & 4.29 & 2.63 & 7.04 & 1.28\\
Multi-task DNN (MSE) + odd--even & $^{235}$U &14 & 12.11 & 28.53 & 7.10 & 2.19 & 5.21 & 1.27\\
Multi-task DNN (Weighted loss function) + odd--even & $^{235}$U &14 & \textbf{3.44} & \textbf{8.32} & \textbf{1.95} & \textbf{1.78} & \textbf{3.56 }& 1.23\\ \hline
\end{tabular}
\end{minipage}

\vspace{4mm}
%\end{table*}
%\end{landscape}
%\hspace{1\linewidth}
%ĂĽÂÂĂ¨ÂĄÂ¨ĂŚÂ Âź
%\begin{landscape}
%\begin{table*}[t]%The best place to locate the table environment is directly after its first reference in the text
\caption{Validation errors $\chi^{2} (10^{-6})$ for $^{238}$U obtained using various methods}
\begin{minipage}{0.40\linewidth}  
\footnotesize
\label{tab:3}
\begin{tabular}{lcccccccc}
&
&
&
\multicolumn{1}{l}{\textrm{total}}&
\multicolumn{1}{l}{\textrm{peak}}&
\multicolumn{1}{l}{\textrm{non-peak}}&
%\multicolumn{3}{c}{\textrm{uncertainty}}\\
\multicolumn{3}{c}{\textrm{FPY error}}\\
% \multicolumn{1}{l}{\textrm{uncertainty}}&
% \multicolumn{1}{l}{\textrm{uncertainty}}\\
\textrm{Methods}&
\textrm{nuclei}&
\textrm{energy}&
\textrm{$\chi^{2}$}&
\textrm{$\chi^{2}$}&
\textrm{$\chi^{2}$}&
% \multicolumn{1}{l}{\textrm{$\chi^{2}$}}&
% \multicolumn{1}{l}{\textrm{$\chi^{2}$}}&
\textrm{total $\chi^{2}$}&
\textrm{peak $\chi^{2}$}&
\multicolumn{1}{l}{\textrm{non-peak $\chi^{2}$}}\\
\hline
%U238
BNN & $^{238}$U &0.5 & 11.33 & 24.06 & 7.45 & 9.99 & 35.94 & 2.08\\
DNN (MSE) & $^{238}$U & 0.5 & 9.98 & 21.09 & 6.60 & 8.01 & 11.28 & 7.02\\
DNN (Weighted loss function) & $^{238}$U & 0.5 & 15.87 & 21.16 & 14.26 & 8.01 & 11.28 & 7.02\\
DNN (MSE) + odd--even & $^{238}$U & 0.5 & 7.45 & 23.84 & 2.46 & 4.92 & 10.97 & 3.07\\
DNN (Weighted loss function) + odd--even & $^{238}$U & 0.5 & 8.80 & 19.75 & 5.46 & 4.92 & 10.97 & 3.07\\
Multi-task DNN (MSE) & $^{238}$U &0.5& 10.48 & 30.67 & \textbf{4.33} & 7.27 & 23.28 & 2.39\\
Multi-task DNN (Weighted loss function) & $^{238}$U &0.5& 9.78 & 22.01 & 6.06 & 9.48 & 29.66 & 3.33\\
Multi-task DNN (MSE) + odd--even & $^{238}$U &0.5 & 9.21 & 24.84 & 4.45 & 2.55 & 8.43 & 0.76\\
Multi-task DNN (Weighted loss function) + odd--even & $^{238}$U &0.5& \textbf{8.53} & \textbf{16.92} & 5.97 & \textbf{1.82} & \textbf{5.71} & \textbf{0.64}\\
\hline
BNN & $^{238}$U &14 & 15.48 & 36.42 & 9.10 & 10.75 & 20.30 & 7.83\\
DNN (MSE) & $^{238}$U & 14 & 15.65 & 31.19 & 10.91 & 7.61 & 11.83 & 6.32\\
DNN (Weighted loss function) & $^{238}$U & 14 & 18.03 & 27.12 & 15.26 & 7.61 & 11.83 & 6.32\\
DNN (MSE) + odd--even & $^{238}$U & 14 & 13.37 & 26.96 & 9.23 & 9.29 & 13.96 & 7.87\\
DNN (Weighted loss function) + odd--even & $^{238}$U & 14 & 11.48 & 22.80 & 8.02 & 9.29 & 13.96 & 7.87\\
Multi-task DNN (MSE) & $^{238}$U &14& 12.79 & 26.88 & 8.50 & 6.03 & \textbf{4.27} & 6.57\\
Multi-task DNN (Weighted loss function) & $^{238}$U &14& \textbf{6.41} & \textbf{14.47} & \textbf{3.95} & \textbf{2.19} & 4.54 & 1.47\\
Multi-task DNN (MSE) + odd--even & $^{238}$U &14& 11.19 & 25.96 & 6.69 & 2.63 & 7.58 & \textbf{1.12}\\
Multi-task DNN (Weighted loss function) + odd--even & $^{238}$U &14& 13.53 & 19.23 & 11.79 & 3.44 & 6.21 & 2.60\\ \hline
\end{tabular}
\end{minipage}
\end{table*}
\end{landscape}

%ĂĽÂÂĂ¨ÂĄÂ¨ĂŚÂ Âź
\begin{landscape}
\begin{table*}[t]%The best place to locate the table environment is directly after its first reference in the text
\caption{Validation errors $\chi^{2} (10^{-6})$ for $^{239}$Pu obtained using various methods } 
\begin{minipage}{0.4\linewidth}
\footnotesize
\label{tab:4}
\begin{tabular}{lcccccccc}
&
&
&
\multicolumn{1}{l}{\textrm{total}}&
\multicolumn{1}{l}{\textrm{peak}}&
\multicolumn{1}{l}{\textrm{non-peak}}&
%\multicolumn{3}{c}{\textrm{uncertainty}}\\
\multicolumn{3}{c}{\textrm{FPY error}}\\
% \multicolumn{1}{l}{\textrm{uncertainty}}&
% \multicolumn{1}{l}{\textrm{uncertainty}}\\
\textrm{Methods}&
\textrm{nuclei}&
\textrm{energy}&
\textrm{$\chi^{2}$}&
\textrm{$\chi^{2}$}&
\textrm{$\chi^{2}$}&
% \multicolumn{1}{l}{\textrm{$\chi^{2}$}}&
% \multicolumn{1}{l}{\textrm{$\chi^{2}$}}&
\textrm{total $\chi^{2}$}&
\textrm{peak $\chi^{2}$}&
\multicolumn{1}{l}{\textrm{non-peak $\chi^{2}$}}\\
\hline

%Pu239
BNN & $^{239}$Pu &0.5 & 5.58 & 7.99 & 4.84 & 10.78 & 36.64 & 2.89\\
DNN (MSE) & $^{239}$Pu & 0.5 & 6.65 & 12.35 & 4.92 & 3.62 & 6.93 & 2.61\\
DNN (Weighted loss function) & $^{239}$Pu & 0.5 & 4.30 & 8.83 & 2.91 & 3.62 & 6.93 & 2.61\\
DNN (MSE) + oddeven & $^{239}$Pu & 0.5 & 4.41 & 8.11 & 3.28 & 4.35 & \textbf{2.95} & 4.77\\
DNN (Weighted loss function) + oddeven & $^{239}$Pu & 0.5 & 4.22 & 7.74 & 3.14 & 4.35 & \textbf{2.95} & 4.77\\
Multi-task DNN (MSE) & $^{239}$Pu &0.5 & 5.13 & 7.65 & 4.36 & \textbf{3.37} & 9.49 & \textbf{0.15}\\
Multi-task DNN (Weighted loss function) & $^{239}$Pu &0.5 & \textbf{2.82} & \textbf{7.13} & \textbf{1.50} & 3.87 & 10.91 & 1.73\\
Multi-task DNN (MSE) + oddeven & $^{239}$Pu &0.5 & 4.37 & 7.60 & 3.39 & 4.02 & 12.01 & 1.58\\
Multi-task DNN (Weighted loss function) + oddeven & $^{239}$Pu &0.5 & 4.25 & 7.57 & 3.24 & 3.91 & 11.90 & 1.47\\
\hline
BNN & $^{239}$Pu &14 & 8.91 & 20.32 & 5.44 &9.71 &18.12 & 7.15\\
DNN (MSE) & $^{239}$Pu & 14 & 8.75 & 21.31 &4.91 & 4.56 & 5.91 & 4.15\\
DNN (Weighted loss function) & $^{239}$Pu & 14 & 9.58 & 20.02 & 6.40 & 4.56 & 5.91 & 4.15\\
DNN (MSE) + oddeven & $^{239}$Pu & 14 & 7.38 & 13.02 & 5.66 & 3.44 & 5.86 & 2.70\\
DNN (Weighted loss function) + oddeven & $^{239}$Pu & 14 & 7.08 & 12.12 & 5.54 & 3.44 & 5.86 & 2.70\\
Multi-task DNN (MSE) & $^{239}$Pu &14 & 7.00 & 16.2 & 4.20 & 2.09 & 4.86 & 1.24\\
Multi-task DNN (Weighted loss function) & $^{239}$Pu &14 & 11.82 & 36.41 & 4.32 & 2.09 & 6.07 & 8.70\\
Multi-task DNN (MSE) + oddeven & $^{239}$Pu &14 & 6.00 & 15.31 & 3.17 & \textbf{1.53} & \textbf{4.50} & \textbf{0.63}\\
Multi-task DNN (Weighted loss function) + oddeven & $^{239}$Pu &14 & \textbf{5.07} & \textbf{11.68} & \textbf{3.07} & 2.23 & 7.01 & 0.77\\ \hline
\end{tabular}
\end{minipage}
%\end{table*}
%\end{landscape}

%ĂĽÂÂĂ¨ÂĄÂ¨ĂŚÂ Âź
\vspace{4mm}

%\begin{landscape}
%\begin{table*}[!htbp]%The best place to locate the table environment is directly after its first reference in the text
\caption{Validation errors $\chi^{2} (10^{-6})$ for $^{241}$Pu obtained using various methods }
\begin{minipage}{0.4\linewidth}
\footnotesize
\label{tab:5}
\begin{tabular}{lcccccccc}
&
&
&
\multicolumn{1}{l}{\textrm{total}}&
\multicolumn{1}{l}{\textrm{peak}}&
\multicolumn{1}{l}{\textrm{non-peak}}&
%\multicolumn{3}{c}{\textrm{uncertainty}}\\
\multicolumn{3}{c}{\textrm{FPY error}}\\
% \multicolumn{1}{l}{\textrm{uncertainty}}&
% \multicolumn{1}{l}{\textrm{uncertainty}}\\
\textrm{Methods}&
\textrm{nuclei}&
\textrm{energy}&
\textrm{$\chi^{2}$}&
\textrm{$\chi^{2}$}&
\textrm{$\chi^{2}$}&
% \multicolumn{1}{l}{\textrm{$\chi^{2}$}}&
% \multicolumn{1}{l}{\textrm{$\chi^{2}$}}&
\textrm{total $\chi^{2}$}&
\textrm{peak $\chi^{2}$}&
\multicolumn{1}{l}{\textrm{non-peak $\chi^{2}$}}\\
\hline
%Pu241
BNN & $^{241}$Pu &0.5 & 3.73 & 6.46 & 2.90 & 9.46 & 29.27 & 3.42\\
DNN (MSE) & $^{241}$Pu & 0.5 & 5.60 & 8.20 & 4.81 & 4.15 & 13.01 & 1.44\\
DNN (Weighted loss function) & $^{241}$Pu & 0.5 & 4.47 & 7.84 & 3.44 & 4.15 & 13.01 & 1.44\\
DNN (MSE) + oddeven & $^{241}$Pu & 0.5 & 4.18 & 7.77 & 3.09 & 3.61 & 11.39 & \textbf{1.25}\\
DNN (Weighted loss function) + oddeven & $^{241}$Pu & 0.5 & 4.14 & 6.80 & 3.32 & 3.61 & 11.39 & \textbf{1.25}\\
Multi-task DNN (MSE) & $^{241}$Pu &0.5 & 4.83 & 14.42 & \textbf{1.90} & \textbf{2.89}& \textbf{7.04} & 1.63\\
Multi-task DNN (Weighted loss function) & $^{241}$Pu &0.5 & 6.52 & 10.09 & 5.43 & 3.34 & 8.47 & 1.77\\
Multi-task DNN (MSE) + oddeven & $^{241}$Pu &0.5 & 5.53 & 8.16 & 4.72 & 3.28 & 7.85 & 1.89\\
Multi-task DNN (Weighted loss function) + oddeven & $^{241}$Pu &0.5 & \textbf{3.49} & \textbf{5.31} & 2.93 & 3.06 & 7.68 & 1.65\\
\hline
% Pu241 0.00025
BNN & $^{241}$Pu &$2.53 \times 10^{-8}$ & 4.66 & 6.74 & 4.03 & 10.45 & 33.79 & 3.34\\
DNN (MSE) & $^{241}$Pu & $2.53 \times 10^{-8}$ & 7.09 & 8.30 & 6.73 & 7.47 & 19.14 & 3.91\\
DNN (Weighted loss function) & $^{241}$Pu & $2.53 \times 10^{-8}$ & 9.51 & 7.18 & 10.22 & 7.47 & 19.14 & 3.91\\
DNN (MSE) + oddeven & $^{241}$Pu & $2.53 \times 10^{-8}$ & 4.58 & 9.36 & 3.13 & 6.38 & 18.59 & 2.65\\
DNN (Weighted loss function) + oddeven & $^{241}$Pu & $2.53 \times 10^{-8}$ & 4.52 & 9.26 & 3.08 & 6.38 & 18.59 & 2.65\\
Multi-task DNN (MSE) & $^{241}$Pu &$2.53 \times 10^{-8}$ & 4.46 & 6.28 & 3.90 & 6.81 & 17.47 & 3.56\\
Multi-task DNN (Weighted loss function) & $^{241}$Pu &$2.53 \times 10^{-8}$ & 5.00 & 5.31 & 4.90 & 7.05 & 18.82 & 3.46\\
Multi-task DNN (MSE) + oddeven & $^{241}$Pu &$2.53 \times 10^{-8}$ & 4.43 & 5.53 & 5.87 & 4.20 & 11.36 & 2.01\\
Multi-task DNN (Weighted loss function) + oddeven & $^{241}$Pu &$2.53 \times 10^{-8}$ & \textbf{3.36} & \textbf{4.26} & \textbf{3.08} & \textbf{3.75} & \textbf{9.87} & \textbf{1.89}\\ \hline

\end{tabular}
\end{minipage}
\end{table*}
\end{landscape}

It is apparent from \autoref{tab:1} that at either energy ($2.53 \times 10^{-8}$ MeV, 0.5 MeV or 14 MeV), the Multi-task DNN with the weighted loss function and the odd--even effect provides the best FPY prediction accuracies for average total $\chi^2$, with the lowest average prediction errors: better than BNN and DNN. 
Although the average errors (total errors $\chi^2$) for the FPY error prediction of the Multi-task DNNs also exhibit the lowest values, the optimal loss function (MSE or Weighted loss function) depends on the nucleus and its excitation energy. The reason is that the loss function weight $\alpha$ in (4) was optimized to maximize only the FPY prediction. 
In addition, the average errors (total errors $\chi^2$) for FPY error prediction of the Multi-task DNN (Weighted loss function)+odd--even model is only slightly inferior to those of the best prediction model.
Consequently, when we specifically examine FPY prediction, the experiment results suggest that the Multi-task DNN (Weighted loss function) + odd--even model is recommended to predict unknown FPY values.

We present all error values in \autoref{tab:2}--\autoref{tab:5} to clarify the superiority of the proposed methods.
In \autoref{tab:2}--\autoref{tab:5}, the term "nuclei" denotes atomic nuclei. "Energy" denotes the excitation energy in units of mega electron volts. 
The terms "total $\chi^2$", "peak $\chi^2$", "non-peak $\chi^2$", "FPY error total $\chi^2$", "FPY error peak $\chi^2$", and "FPY error non-peak $\chi^2$" are the same as those in \autoref{tab:1}.
The definitions of "total $\chi^2$", "peak $\chi^2$", "non-peak $\chi^2$", "FPY error total $\chi^2$", "FPY error peak $\chi^2$", and "FPY error non-peak $\chi^2$" remain consistent those of earlier descriptions.

%%%%%%%%%%%%%%%%%%%%%%%%%%%%%%%%%%ĂŚÂÂšĂĽÂÂ°Ă¨ÂżÂE

Findings presented in \autoref{tab:2}--\autoref{tab:5} demonstrate that the Multi-task DNN methods provide lower $\chi^{2}$ values for predictions of FPY compared to BNN and DNN in most cases. The reason is that MMoE for the simultaneous learning of both FPY and FPY error improves the prediction accuracies because the FPY error value is related to the FPY value. 
They affect one another, which increases their learning accuracies.

\autoref{tab:2} and \autoref{tab:3} signify that the DNN with MSE loss function provides lower values of both total $\chi^{2}$ and peak $\chi^{2}$ than BNN does for $^{235}$U and $^{238}$U at both 0.5 MeV and 14 MeV. 
However, for $^{239}$Pu at 0.5 MeV in \autoref{tab:4}, as well as $^{241}$Pu at both 0.5 MeV and $2.53 \times 10^{-8}$ MeV in \autoref{tab:5}, the DNN with MSE loss function provides higher values of both total $\chi^{2}$ and peak $\chi^{2}$ compared to the BNN.

For prediction of the FPY error, except for $^{235}$U at 14 MeV, the FPY error total $\chi^{2}$ and the FPY error peak $\chi^{2}$ values of the DNN are lower than those of the BNN.

In summary, the DNN demonstrates comparable predictive capability to the BNN for FPY. In contrast, it demonstrates an advantage over the BNN in predicting FPY error, achieving consistently higher accuracy.

Furthermore, \autoref{tab:2}--\autoref{tab:5} indicate that the Multi-task DNN with MSE loss function provides lower values of total $\chi^{2}$ than DNN with the MSE loss function does for $^{239}$Pu at either 0.5 MeV or 14 MeV, $^{241}$Pu at 0.5 MeV or $2.53 \times 10^{-8}$ MeV, or for $^{238}$U at 14 MeV. Although the $\chi^{2}$ values for $^{235}$U (0.5 MeV and 14 MeV) and $^{238}$U (0.5 MeV) are higher for the Multi-task DNN with the MSE loss function compared to the DNN with the MSE loss function, they are still lower than those of the BNN.

For the prediction of FPY error, the values of FPY error total $\chi^{2}$ from the Multi-task DNN with the MSE loss function are lower than those of the DNN with the MSE loss function, except for $^{238}$U at 0.5 MeV.

% By contrast, no significant differences were found between the $\chi^{2}$ values of the Multi-task DNN and those of the BNN for $^{238}$U, $^{239}$Pu, and $^{241}$Pu at 0.5 MeV, and for $^{241}$Pu at $2.53 \times 10^{-8}$ MeV.

% \autoref{tab:2} signifies that the Multi-task DNN with MSE loss function provides lower values of both $\chi^{2}$ than BNN does for $^{235}$U at both 0.5 MeV and 14 MeV, as well as for $^{238}$U and $^{239}$Pu at 14 MeV. 
% By contrast, no significant differences were found between the $\chi^{2}$ values of the Multi-task DNN and those of the BNN for $^{238}$U, $^{239}$Pu, and $^{241}$Pu at 0.5 MeV, and for $^{241}$Pu at $2.53 \times 10^{-8}$ MeV.

%ĂŚÂÂ Ă¨ÂŽÂşFPYĂ¨ÂżÂĂŚÂÂŻErrorÂEÂDNNĂĽÂ°ÂĂ¤ÂşÂBNN
% Next, we validate the effectiveness of the weighted loss function by comparing the error values of "Multi-task DNN(MSE)" and "Multi-task DNN(Weighted loss function)".

Additionally, differences in FPY error predictions among the Multi-task DNN methods are slight because the loss functions for the FPY error prediction task of Multi-task DNN are the same. 
It is noteworthy that negative predictions of FPY occur frequently in BNN methods, with a proportion of 17.52\%. 
By contrast, Multi-task DNN methods demonstrate high predictive accuracy, producing negative predictions of FPY at a lower rate of 8.29\%. 
Even without additional measures, such as incorporating extra data or modifying activation functions, negative predictions in Multi-task DNN methods are fewer than those obtained using BNN methods.

\subsubsection{Performance comparison of the proposed method with the weighted loss function}
In this subsection, 
we validate the weighted loss function effectiveness by comparing the error values obtained for "Multi-task DNN (MSE)" and "Multi-task DNN (Weighted loss function)" and between "DNN (MSE)" and "DNN (Weighted loss function)".

For DNN, \autoref{tab:2}--\autoref{tab:5} show that, except for $^{238}$U at 0.5 MeV, the peak $\chi^{2}$ values of DNN with the weighted loss function are lower than those of DNN with MSE across all energy levels (0.5 MeV, 14 MeV, and $2.53 \times 10^{-8}$ MeV).

For Multi-task DNN, \autoref{tab:2}--\autoref{tab:5} indicate that, in the case of 0.5 MeV, the peak $\chi^{2}$ values of Multi-task DNN with the weighted loss function are lower than those of Multi-task DNN with MSE for $^{235}$U, $^{238}$U, $^{239}$Pu and $^{241}$Pu, particularly for $^{238}$U. This result represents marked improvement in the peak prediction accuracy using the weighted loss function. Additionally, except for $^{241}$Pu, the total $\chi^{2}$ values for $^{235}$U, $^{238}$U, $^{239}$Pu are lower than those of Multi-task DNN with MSE.
For FPY error prediction, the performances of the weighted loss function and MSE are not markedly different.

Findings presented in \autoref{tab:2}--\autoref{tab:5} signify that, in the cases of 14 and $2.53 \times 10^{-8}$ MeV, the peak $\chi^{2}$ values of Multi-task DNN with the weighted loss function are lower than those of Multi-task DNN with MSE for $^{235}$U, $^{238}$U and $^{241}$Pu.
However, both the peak $\chi^{2}$ value and the total $\chi^{2}$ value of the Multi-task DNN with the weighted loss function are higher than those of the Multi-task DNN with MSE for $^{239}$Pu.
For FPY error prediction, similarly to the results obtained for 0.5 MeV, total $\chi^{2}$ values with the weighted loss function are higher than those with MSE, except for $^{238}$U.

\subsubsection{Performance comparison of the proposed method with the odd–even effect}
In this subsection, 
Finally, 
we demonstrate that the odd--even effect parity indicator $O_i$ is effective for enhancing the FPY prediction accuracy.

For DNN, \autoref{tab:2}--\autoref{tab:5} show that, except for the cases of $^{235}$U at 0.5 MeV (MSE loss function) and 14 MeV (weighted loss function), the total $\chi^{2}$ values of methods with the odd--even effect are consistently lower than those without the odd--even effect.

For Multi-task DNN, at 0.5 MeV for $^{235}$U, $^{238}$U, and $^{241}$Pu, at 14 MeV for $^{238}$U and $^{239}$Pu, and at $2.53 \times 10^{-8}$ MeV for $^{241}$Pu, the methods with the odd--even effect consistently exhibit lower peak $\chi^{2}$ and total $\chi^{2}$ values compared to those without the odd--even effect, irrespective of whether the loss function is the weighted loss function or MSE. This finding signifies that introduction of the odd--even effect is effective in most cases.

However, for $^{239}$Pu at 0.5 MeV and $^{238}$U at 14 MeV, the Multi-task DNNs with the weighted loss function and the odd--even effect exhibit higher peak $\chi^{2}$ and total $\chi^{2}$ values than the Multi-task DNNs with the weighted loss function but without the odd--even effect. The reason for that difference is that the peak structures of the FPY are smooth. Details underlying the reason for that finding are presented later using figures for the prediction of FPY.

%ĺçEĺĺ ä¸čŹçŻEšEźĺć E
% \begin{figure*}[htbp]
%     \centering
%     \begin{minipage}[t]{0.48\textwidth}
%         \centering
%         \includegraphics[width=0.80\textwidth,height=0.55\textheight]{compare0.5_last.pdf}
%         \caption{FPY prediction of $^{235}U$ at 0.5 MeV using various methods.}
%         \label{fig:2}
%     \end{minipage}
%     \hfill
%     \begin{minipage}[t]{0.48\textwidth}
%         \centering
%         \includegraphics[width=0.80\textwidth,height=0.55\textheight]{compare_14Mev_lr10-4_last.pdf} % ĂÂŠĂÂĂÂĂEĂEźĂÂEwidth ĂÂŠĂÂĂÂĂÂŠĂEEÂĄĂEĂE˝ĂEžĂÂ§ĂÂĂE˝ĂEşĂÂĂEžĂEşĂÂĂÂĂEşĂE˝ĂÂĂEşĂEĂEE
%         \caption{FPY prediction of $^{235}$U at 14 MeV using various methods.}
%         \label{fig:3}
%     \end{minipage}
% \end{figure*}

Similarly, from \autoref{tab:2} to \autoref{tab:5}, it is readily apparent that DNNs with the odd--even effect do not consistently exhibit higher or lower FPY error total $\chi^{2}$ values and FPY error peak $\chi^{2}$ values compared to those without the odd--even effect. 
This result is consistent with the fact that the literature has confirmed clear correlation between the odd--even effect and FPY error data \cite{25schmidt2018review}.

\subsubsection{Visual performance comparison among BNN, DNN and the proposed method}
\begin{figure*}[!htbp]
    \centering
    \includegraphics[width=1\textwidth,height=0.8\textheight]{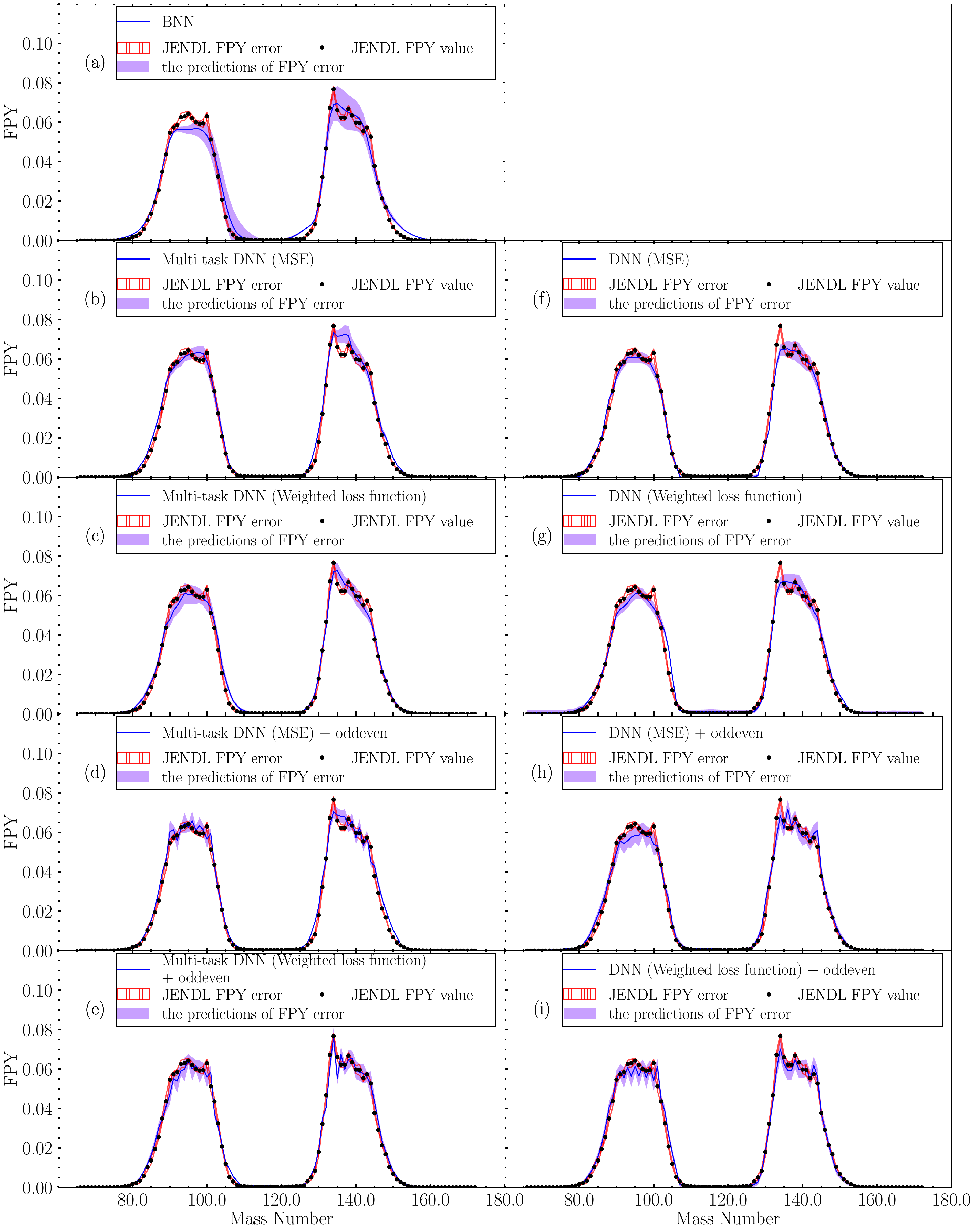}
    \caption{FPY prediction of $^{235}$U at 0.5 MeV using various methods.}
    \label{fig:2}
\end{figure*}
%Additionally, differences in FPY error predictions among the Multi-task DNN methods are slight because the loss functions for the FPY error prediction task of Multi-task DNN are the same.

%It is noteworthy that negative predictions of FPY occur frequently in BNN methods, with a proportion of 17.52\%. 
%By contrast, Multi-task DNN methods demonstrate high predictive accuracy, producing negative predictions of FPY at a lower rate of 8.29\%. 
%Even without additional measures, such as incorporating extra data or modifying activation functions, negative predictions in Multi-task DNN methods are fewer than those obtained using BNN methods.
% Importantly, it is noteworthy that across both BNN and Multi-task DNN methods, an occasional occurrence of negative predictions of FPY is observed within the non-peak data. 

% To address this issue, incorporating experimental data or modifying the activation functions within the model could prevent these unrealistic negative values. }

\begin{figure*}[!htbp]
    \centering
    \includegraphics[width=1\textwidth,height=0.8\textheight]{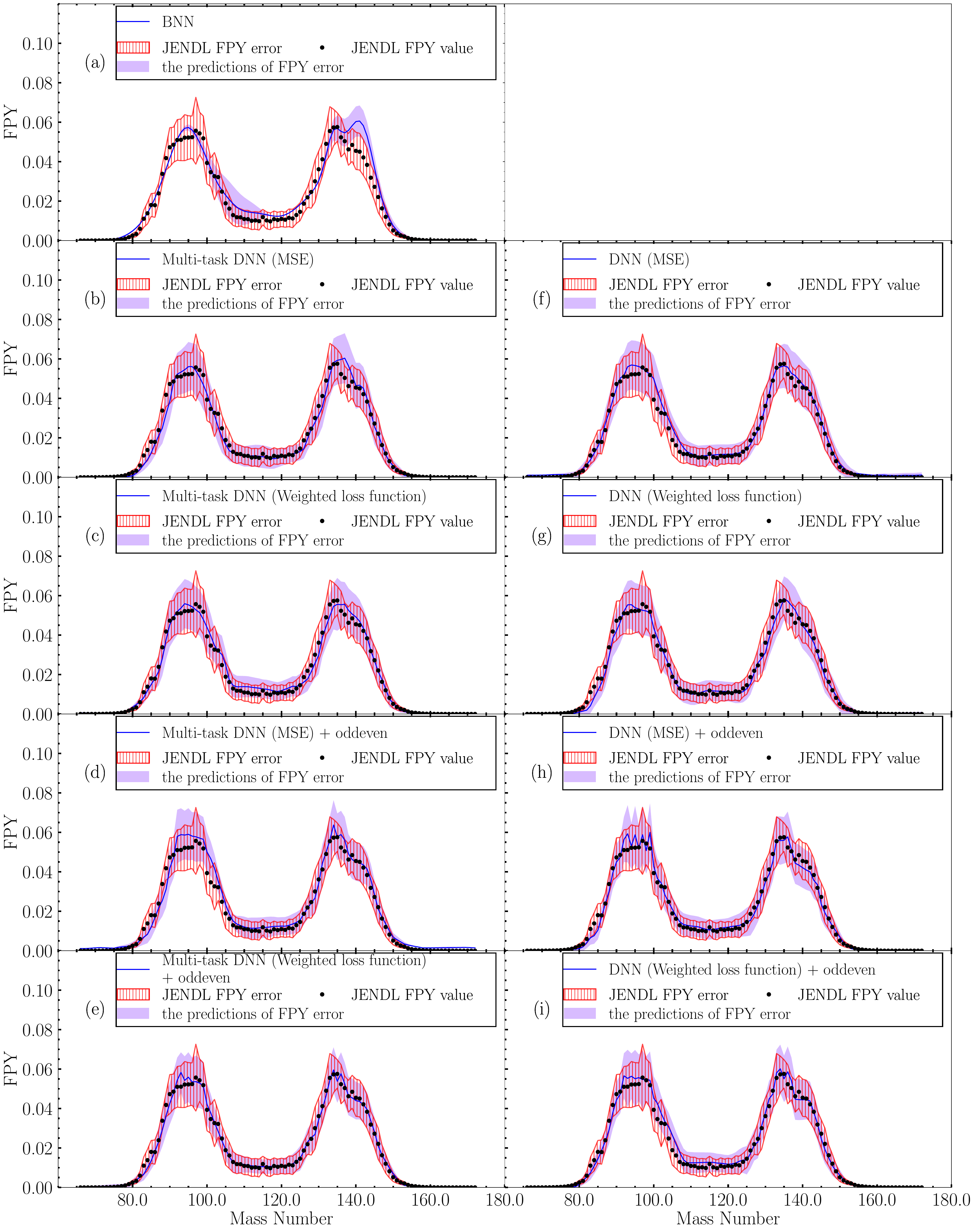}
    \caption{FPY prediction of $^{235}$U at 14 MeV using various methods.}
    \label{fig:3}
\end{figure*}

\begin{figure*}[!htbp]
    \centering
    \begin{minipage}[b]{0.48\textwidth}
        \centering
        \includegraphics[width=1\textwidth,height=0.35\textheight]{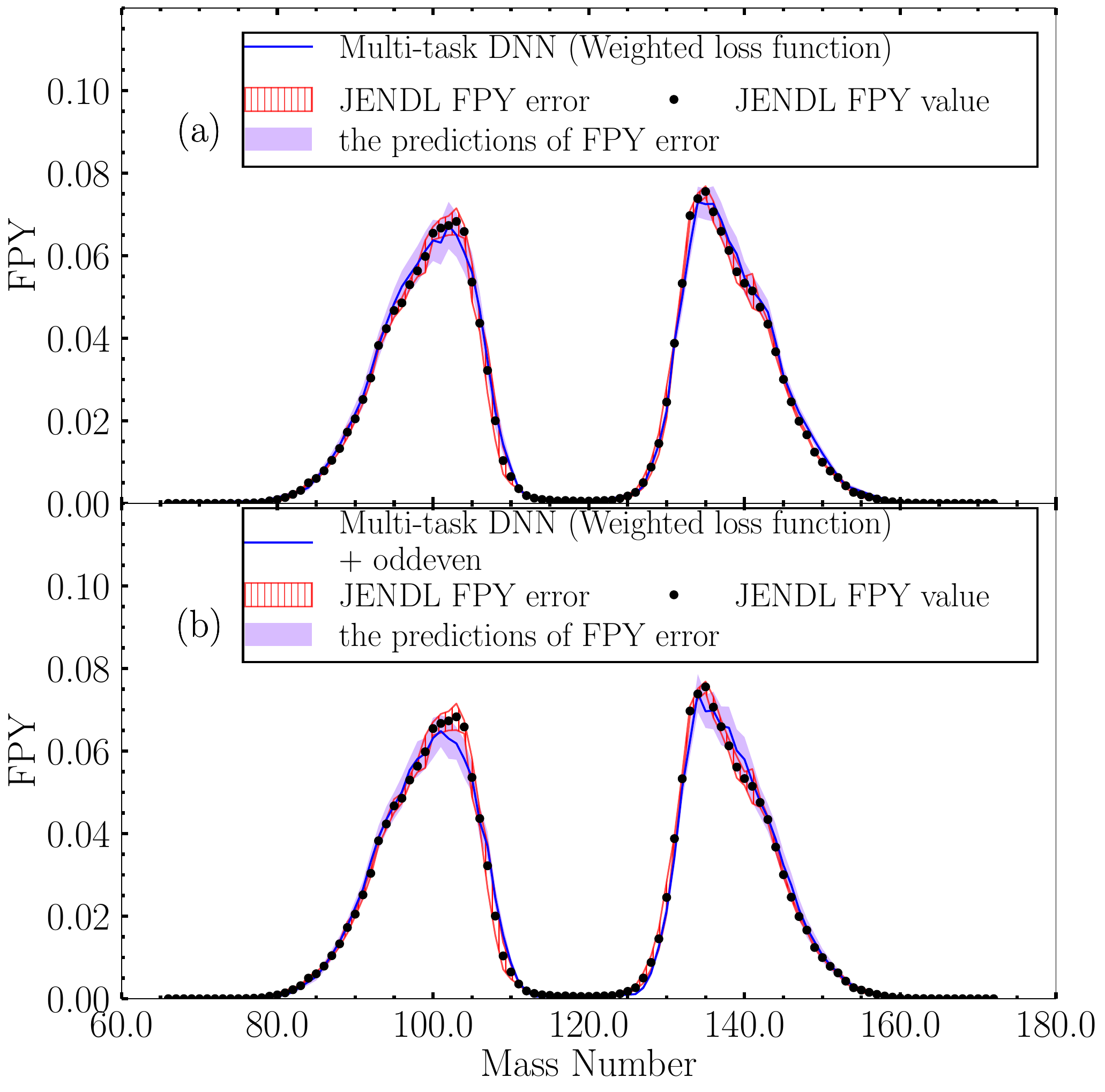} 
        \caption{FPY prediction of $^{239}$Pu at 0.5 MeV using a weighted loss function and odd--even.}
        \label{fig:4}
    \end{minipage}
    \hfill
    \begin{minipage}[b]{0.48\textwidth}
        \centering
        \includegraphics[width=1\textwidth,height=0.35\textheight]{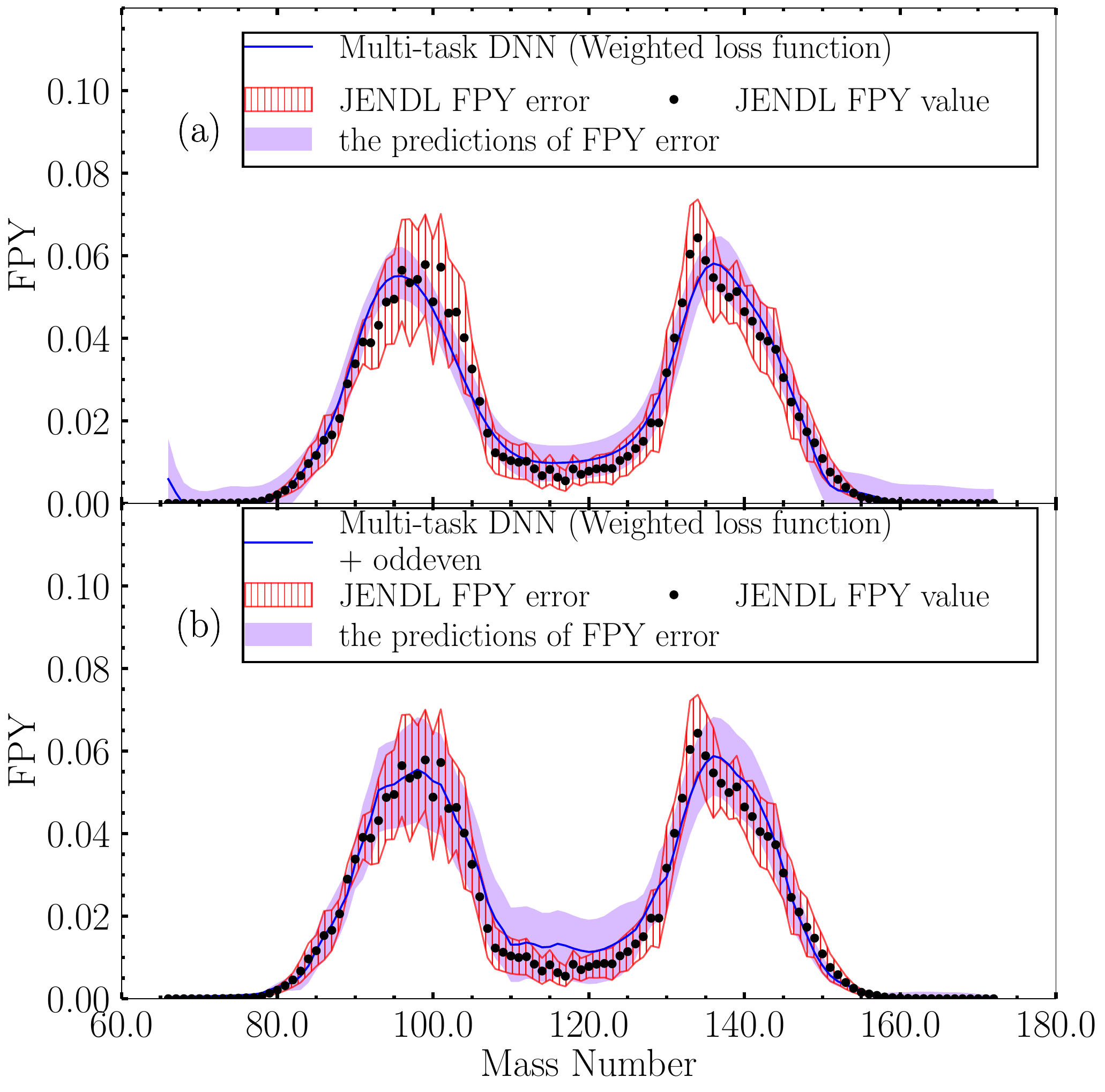} % ĂÂŠĂÂĂÂĂEĂEźĂÂEwidth ĂÂŠĂÂĂÂĂÂŠĂEEÂĄĂEĂE˝ĂEžĂÂ§ĂÂĂE˝ĂEşĂÂĂEžĂEşĂÂĂÂĂEşĂE˝ĂÂĂEşĂEĂEE
        \caption{ FPY prediction of $^{238}$U at 14 MeV using a weighted loss function and odd--even.}
        \label{fig:5}
    \end{minipage}
\end{figure*}

%For visual demonstration of the effectiveness of our proposed methods in predicting FPY and FPY error, 
\autoref{fig:2} and \autoref{fig:3} respectively depict the predictions of FPY for $^{235}$U at 0.5 MeV and 14 MeV using BNN, DNN and the proposed method.
%because our proposed method with the weighted loss function and the odd--even effect provides the highest precision for these predictions.

In the respective figures, the horizontal axes show the mass numbers, whereas the vertical axes show the FPY values. 
Black dots represent the true values of FPY data from JENDL-5 ("JENDL FPY value"). 
Blue lines represent the predictions of the FPY values using "BNN", "DNN" and "Multi-task DNN".
The purple shaded regions represent the predictions of FPY errors using "BNN", "DNN" and "Multi-task DNN".
The red shaded regions represent the true value of FPY error data from JENDL-5 ("JENDL FPY error").

%The compositions of the figures are designed to demonstrate the effectiveness of each method: 
%\autoref{fig:2}(a) and \autoref{fig:3}(a) depict the predictions of FPY and FPY error using BNN.
%\autoref{fig:2}(b) and \autoref{fig:3}(b) depict the predictions of FPY and FPY error using Multi-task DNN with the MSE loss function.
%\autoref{fig:2}(c) and \autoref{fig:3}(c) depict the predictions of FPY and FPY error using Multi-task DNN with the weighted loss function.
%\autoref{fig:2}(d) and \autoref{fig:3}(d) depict the predictions of FPY and FPY error using Multi-task DNN with the MSE loss function and odd--even effect.
%\autoref{fig:2}(e) and \autoref{fig:3}(e) depict the predictions of FPY and FPY error using Multi-task DNN with the weighted loss function and odd--even effect.
%\autoref{fig:2}(f) and \autoref{fig:3}(f) depict the predictions of FPY and FPY error using DNN with the MSE loss function.
%\autoref{fig:2}(g) and \autoref{fig:3}(g) depict the predictions of FPY and FPY error using DNN with the weighted loss function.
%\autoref{fig:2}(h) and \autoref{fig:3}(h) depict the predictions of FPY and FPY error using DNN with the MSE loss function and odd--even effect.
%\autoref{fig:2}(i) and \autoref{fig:3}(i) depict the predictions of FPY and FPY error using DNN with the weighted loss function and odd--even effect.

\autoref{fig:2} demonstrates that the predictions of FPY for $^{235}$U obtained by the Multi-task DNN methods in \autoref{fig:2}(b)--(e) are closer to the true FPY values from JENDL-5 than the BNN results shown in \autoref{fig:2}(a), particularly in the peak region. 
The proposed Multi-task DNN improves predictive accuracy because it directly and simultaneously learns the evaluated nuclear data FPY error together with FPY value, exploiting their complementary relationship during optimization.
Among these Multi-task DNN methods, \autoref{fig:2}(e), which incorporates both the weighted loss function and the odd--even effect, provides the to the true value of FPY data from JENDL-5 in all methods.
Specifically, in the actinoid region, the Multi-task DNN method using the weighted loss function and the odd--even effect reproduces the characteristic double-humped FPY structure influenced by the spherical magic number at $A=132$ and the deformed magic number at $A=142$, with fine oscillatory structures arising from the odd--even effect.

\autoref{fig:3} presents that, even at 14 MeV, the Multi-task DNN with the weighted loss function and the odd--even effect still exhibits the closest predictions of FPY to the true value of FPY data from JENDL-5. 
However, for all of these methods, the FPY error predictions are larger in the case of 14 MeV than for 0.5 MeV because the covariance matrix evaluated by Tsubakihara et al. \cite{FPY_covariance} is not available at 14 MeV. In this case, the FPY error is estimated as the square root of the sum of squared individual errors, resulting in larger predicted errors.
By contrast, in \autoref{fig:2}, the FPY error for 0.5 MeV is estimated using the covariance matrix, resulting in small predicted errors.

It can be observed that the predictions of FPY using Multi-task DNN with the weighted loss function in \autoref{fig:4}(a) and \autoref{fig:5}(a) are sufficiently close to the true value of FPY data from JENDL-5. 
However, after employing the odd--even effect, the predictions of FPY at peak structures in \autoref{fig:4}(b) and \autoref{fig:5}(b) get farther away from the true value of FPY data from JENDL-5. 
Additionally, one can observe that the shaded region of the predictions of FPY error in \autoref{fig:5}(a) overlaps more with the shaded region of the true value of FPY error data from JENDL-5.
These phenomena might occur because of the extremely smooth peak structures of the FPY for $^{239}$Pu at 0.5 MeV and $^{238}$U at 14 MeV, where the jagged features might not be readily apparent. 
Therefore, the odd--even effect is ineffective at increasing the learning accuracies for these data.

Finally, these predictions from the Multi-task DNN with the weighted loss function and the odd--even effect align closely with known physical principles \cite{27iwamoto2023japanese,28brosa1990nuclear}. The global double-humped structure widens because of the rise in the temperature of the fissioning nucleus, leading to decreased yield at the peaks and increased yield in the tails. Simultaneously, the fine structures become smoother because peak yields in the fine structure decrease while the yields in the valleys increase.
Additionally, the peak of the heavier fission fragment shifts to smaller $A$ values (to the left) because of an increase in the prompt neutron yield from the heavier fragments. When the neutron energy exceeds several mega electron volts and multi-chance fission occurs, the contributions from fissioning nuclei with mass numbers reduced by 1, 2, or more become considerable, causing the peak of the lighter fission fragment to shift further leftwards.
Specifically, the widening of the global double-humped structure and the smoothing of fine structures correspond to the thermal effects and neutron emission dynamics of the fissioning nucleus. The shifts in the peaks of the lighter and heavier fission fragments are also consistent with changes induced by increased prompt neutron production and multi-chance fission. Our results demonstrate that the predictions of the Multi-task DNN with the weighted loss function and the odd--even effect not only reproduce experiment trends but also align closely with these established physical mechanisms.
The predictions of FPY error reflect the reliability of our knowledge about that quantity and serve as a basis for evaluating the theoretical model validity. From an engineering perspective, it provides an estimate of the uncertainty in the amount of radioactive nuclides produced in nuclear reactors, offering crucially important guidance for the design feasibility of reprocessing and geological disposal facilities.

\subsection{Evaluation of FPY at different incident energies}

\begin{figure*}[!b]
    \centering
    \includegraphics[angle=90,width=0.46\textheight]{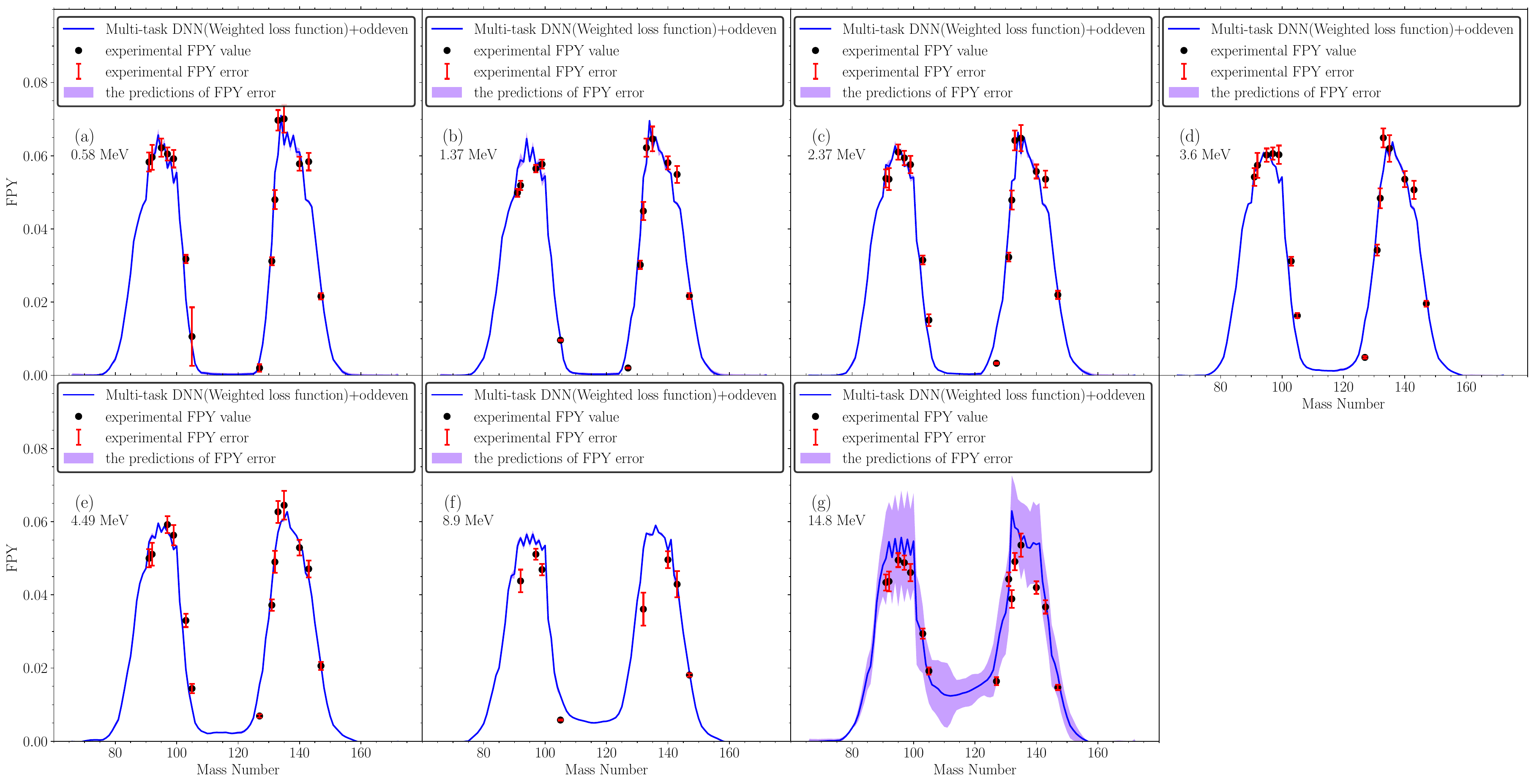}
   \caption{Compiled evaluations of FPY and FPY error for $^{235}$U at different incident energies using Multi-task DNN(Weighted loss function + odd--even).}
    \label{fig:6}
\end{figure*}

Our proposed methods are motivated primarily by the need to predict incomplete experimentally obtained FPY data using information from completed assessments of nuclei.
\autoref{fig:6} presents predictions of FPY and FPY error using the Multi-task DNN with the weighted loss function and the odd--even effect at various energies.

\begin{table*}[t]%The best place to locate the table environment is directly after its first reference in the text
\caption{Validation errors $\chi^{2} (10^{-6})$ of various supplementary datasets for $^{235}$U at 0.5 MeV}
\footnotesize
\label{tab:6}
\begin{tabular}{ccccc}
\textrm{Number of divisions} &
\textrm{quantity} &
\textrm{energy} &
\textrm{total $\chi^{2}$} &
\textrm{peak $\chi^{2}$} \\
\hline
- & 0   & -   & 5.34  & 12.67 \\
2 & 107 & 7.00 & 4.51  & 8.97  \\
3 & 214 & 4.67, 9.33 & 4.27  & 9.59  \\
4 & 321 & 3.50, 7.00, 10.50 & 4.06  & 10.73 \\
5 & 428 & 2.80, 5.60, 8.40, 11.20 & 3.07  & 6.82  \\
6 & 535 & 2.33, 4.67, 7.00, 9.33, 11.67 & \textbf{2.06} & \textbf{6.37} \\
7 & 642 & 2.00, 4.00, 6.00, 8.00, 10.00, 12.00 & 3.26  & 9.71  \\
8 & 749 & 1.75, 3.50, 5.25, 7.00, 8.75, 10.50, 12.25 & 3.42  & 9.93  \\
9 & 856 & 1.56, 3.11, 4.67, 6.22, 7.78, 9.33, 10.89, 12.44 & 4.13  & 9.34 \\ \hline
\end{tabular}
\end{table*}

For this experiment, we expand the training dataset to include $^{235}$U from JENDL-5. 
Additionally, we introduce supplementary data generated from the Gaussian model \cite{26katakura2003systematics} because the training data for the unknown energies are insufficient. 
The Gaussian model represents the fission product mass-yield distribution as a superposition of five Gaussian functions, each corresponding to characteristic components such as the light and heavy peaks and the symmetric component.
The fission product mass-yield distribution provides an analytical approximation of the mass distribution and has been widely used in nuclear data evaluations.
These supplementary data augment the training data and function similarly to a prior distribution of Bayesian inference.
In this study, we employ this model to generate supplementary FPY data at selected excitation energies, thereby compensating for the scarcity of experimental data in the intermediate energy region.
It should be noted that in this study, the Katakura Gaussian model was utilized solely to provide supplementary analytical FPY values for the intermediate energy region, where experimental information is scarce.
Because this phenomenological model lacks corresponding uncertainty information, the FPY error values for these supplementary data were assigned as zero within the training dataset. 
Consequently, no correlations were introduced into these analytical data.

For this experiment, we expand the training dataset to include $^{235}$U from JENDL-5. Additionally, we introduce supplementary data generated from the Gaussian model \cite{26katakura2003systematics} because the training data for the unknown energies are insufficient. These supplementary data augment the training data and function similarly to a prior distribution of Bayesian inference.
Nevertheless, the FPY error values for these supplementary data are set to zero because it is difficult to estimate the FPY error for the supplementary data generated by the Gaussian model.
Although this simplification enables the model to learn the structural features of the mass-yield distribution, it  may lead to an underestimation of FPY error in energy regions where experimental covariance information is unavailable.

% ĂĽÂÂ Ă¤Â¸ÂşĂ¨ÂĄÂĽĂĽÂEEÂÂ°ĂŚÂÂŽĂ§ÂÂEÂÂ°ĂŠÂEĂ¤ÂźÂĂĽÂŻÂšFPYĂŠÂ˘ÂEÂľÂĂŠÂÂ ĂŚÂÂĂĽÂ˝ÂąĂĽÂÂĂŻÂźÂĂŚÂÂĂ¤ÂťÂĽĂŚÂÂĂ¤ÂťÂŹĂĽÂ°ÂE.5MeVĂ§ÂÂU235Ă¤Â˝ÂĂ¤Â¸ÂşĂŚÂľÂĂ¨ÂŻÂĂŠÂÂEŚÂÂĽĂŚÂľÂĂ¨ÂŻÂĂĽÂ¤ÂĂ§ÂťÂEÂĄÂĽĂĽÂEEÂÂ°ĂŚÂÂŽĂ§ÂÂEÂÂĂŚÂÂEĂ¤ÂťÂĽĂŚÂ­Â¤ĂŚÂÂĽĂŠÂÂĂŚÂÂŠĂŚÂÂĂ¤Â˝ÂłĂ§ÂÂEÂĄÂĽĂĽÂEEÂÂ°ĂŚÂÂŽĂŁÂÂE

To identify the supplementary dataset which achieves optimal performance for the FPY prediction, we employ $^{235}$U at 0.5 MeV from JENDL-5 as the test dataset to evaluate the effects of different supplementary datasets on prediction accuracy.

The energy range of 0--14 MeV is divided into multiple parts at equal intervals. The division points are then used as excitation energies to generate supplementary data based on the Gaussian model. Details are provided in \autoref{tab:6}, where the "Number of divisions" represents the number of equal divisions of the 0 to 14 MeV range, "quantity" denotes the amount of supplementary data, and "energy" represents the excitation energies of supplementary data in Me V. The meanings of "total $\chi^2$" and "peak $\chi^2$" are consistent with those in \autoref{tab:1}--\autoref{tab:5}. The case in which the number of divisions is marked as -- indicates that no supplementary data are included in the training dataset. This case is treated as the baseline.

As shown in \autoref{tab:6}, the values of "total $\chi^2$" and "peak $\chi^2$" achieve their minima at six divisions. Consequently, we incorporate the 535 supplementary data points from the six-division scenario into the training dataset to predict incomplete experimental FPY data.

\autoref{fig:6} depicts the predictions of FPY and FPY error using the Multi-task DNN with the weighted loss function and the odd--even effect at various energies: (a) 0.58 MeV, (b) 1.37 MeV, (c) 2.37 MeV, (d) 3.6 MeV, (e) 4.49 MeV, (f) 8.9 MeV, and (g) 14.8 MeV. 
In \autoref{fig:6}, the horizontal axis shows the mass number, whereas the vertical axis shows the FPY value. 
The blue lines represent the predictions of the FPY values. The black dots represent the limited experimental value of FPY data ("experimental FPY value") \cite{peak1gooden2016energy}. 
The red bars represent the limited experimental value of FPY error data ("experimental FPY error") \cite{peak1gooden2016energy}. 
The purple shaded regions represent the predictions of FPY error.

\autoref{fig:6} presents that the Multi-task DNN with the weighted loss function and the odd--even effect exhibits the close predictions of FPY to the experimentally obtained value of FPY data \cite{peak1gooden2016energy}. 
Additionally, the predictions of FPY error in Fig. \autoref{fig:6} are aligned closely with the experimental value of FPY error data \cite{peak1gooden2016energy}.

In \autoref{fig:6}, an observable trend emerges: as the excitation energy increases, the fission yield in the peak region decreases gradually, whereas the valley region (around mass numbers 110--120) exhibits an upward trend. This trend aligns with the energy dependence pattern of $^{235}$U \cite{27iwamoto2023japanese,28brosa1990nuclear}. 
Furthermore, irrespective of variations in the incident energy, the highest point of the right peak structure remains around mass number 134, which is consistent with results obtained from earlier studies \cite{peak1gooden2016energy,peak2tonchev2020toward}. 
These observed features demonstrate the capability of a Multi-task DNN with the weighted loss function and the odd--even effect to learn and predict energy-dependent FPY effectively.

On the other hand, the FPY error values exhibit energy-dependent behavior in \autoref{fig:6}. 
The FPY error values are small in subfigure panels (a)--(f), because these predicted energies are close to those of the supplementary datasets. 
Since the supplementary training data were generated using the Gaussian model and their FPY error values were set to zero, the predicted FPY error values in these panels tend to be small. 
In contrast, subfigure panel (g), corresponding to 14.8 MeV, shows considerably larger FPY error values. 
This is reasonable because 14.8 MeV is closest to 14 MeV in JENDL-5, for which evaluated FPY error values based on the covariance matrix by Tsubakihara et al. \cite{FPY_covariance} are available. 
Therefore, at 14.8 MeV, the Multi-task DNN with the weighted loss function and the odd--even effect reflects the influence of non-zero FPY error training data, whereas at other energies the FPY error predictions are suppressed due to the absence of reliable error information in the supplementary datasets.

\section{\label{sec:SUMMARY}SUMMARY}

We have introduced an application of a Multi-task DNN to learn and predict FPY and FPY error of actinide nuclei. 
Multi-task DNN architecture enables learning of FPY data and learning their FPY errors supplementarily while assisting each other. 
Therefore, the proposed method enhanced the prediction accuracy of FPY.
Simultaneously learning both FPY and FPY error captures their common features more effectively.
This innovative approach incorporates a novel loss function and the odd--even effect to improve the accuracy of predicting peak-shaped FPY data considerably while estimating and evaluating the FPY error. 
Our study has demonstrated the effectiveness of the proposed loss function and the odd--even effect for improving the predictive capabilities of the Multi-task DNN, yielding satisfactory outcomes related to the peak-shaped distribution, predicting FPY data, FPY error data simultaneously, and energy dependence of FPY. 

Our research has highlighted the enhancement of Multi-task DNN learning and predictive abilities for FPY and FPY error to predict the peak-shaped distribution accurately, emphasizing the need for continued refinement of the loss functions. 
As future work, we expect to improve loss functions of the Multi-task DNN approach to improve the peak structure prediction performance.

Because a BNN estimates the Gaussian distribution to approximate the data distribution, predictive evaluation based only on the  expected value estimates of FPY without using CIs might be an impediment to the BNN.
Alternatively, although this study did not use the CIs of the BNN in the experiments, it can be inferred that the BNN has high predictive accuracy if evaluated based on whether the confidence interval includes the true value.
Consequently, it is noteworthy that the comparison results for the proposed method and BNN in this study specifically emphasize prediction of the detailed structure of peaks. 
Accordingly, it is not claimed for this study that the proposed method is superior to BNN for predicting FPY data because the prediction objectives of BNN and those of the proposed method differ.
Using different methods depending on the purpose is important.

%\section{\label{sec:LIMITATION}LIMITATION}
A limitation of the present study is that validation was conducted only for major actinides ($^{235}$U, $^{238}$U, $^{239}$Pu, and $^{241}$Pu), for which abundant and reliable evaluated FPY data are available.
The generalization capability of the proposed method to other nuclides remains to be systematically verified using sufficient experimental datasets.
Future work will extend validation to additional nuclides as more comprehensive experimental data become available, in order to assess the robustness and generalization performance of the proposed multi-task framework.

In addition, the supplementary FPY data generated from the Katakura Gaussian model do not include corresponding FPY error values.
Therefore, the FPY error of the supplementary data was set to zero for training data.
Although this simplification enables the model to learn the structural features of the mass-yield distribution, it may lead to an underestimation of FPY error in energy regions where experimental covariance information is unavailable.

In future work, we plan to establish a simulation-based framework to estimate FPY error for excitation energies where experimental covariance data are insufficient.
Incorporating physically consistent uncertainty estimation into the supplementary dataset will be expected to improve the reliability and predictive robustness of the proposed multi-task model.

\section*{Acknowledgement}
This work was supported by Ministry of Education, Culture, Sports, Science and Technology MEXT Innovative Nuclear Research and Development Program Grant Number JPMXD0222682620, entitled ``Fission product yields predicted by machine learning technique at unmeasured energies and its influence on reactor physics assessment,’’ entrusted to the Tokyo Institute of Technology.
%"Fission product yields predicted using machine learning techniques at unmeasured energies and their influence on reactor physics assessment" was entrusted to the Tokyo Institute of Technology by the Ministry of Education, Culture, Sports, Science and Technology of Japan (MEXT).

%\section*{Funding}
%
%An unnumbered section, e.g.\ \verb"\section*{Funding}", may be used for grant details, etc.\ if required and included \emph{in the non-anonymous version} before any Notes or References.

\bibliographystyle{tfnlm}
\bibliography{yUenoPRCMultiTaskcompareVer}

\end{document}